\documentclass[review,3p]{elsarticle}
\usepackage{hyperref}
\usepackage{graphicx}
\usepackage{subfigure}
\usepackage[table]{xcolor} 
\usepackage{multirow}
\usepackage{tabularx}
\usepackage{siunitx}
\usepackage{booktabs}
\usepackage{times}
\usepackage{CJK}
\usepackage{amsmath}
\usepackage{verbatim}

\journal{Physica A}

\begin{document}

\begin{frontmatter}

\title{Experimental study on the evading behavior of individual pedestrians when confronting with an obstacle in a corridor}

\author[me]{Xiaolu Jia$^*$}
\address[me]{Department of Advanced Interdisciplinary Studies, 
Graduate School of Engineering,\\The University of Tokyo, 4-6-1 Komaba, 
Meguro-ku, Tokyo 153-8904, Japan}

\cortext[mycorrespondingauthor]{Corresponding author}
\ead{xiaolujia@g.ecc.u-tokyo.ac.jp}

\author[RCAST]{Claudio Feliciani}

\address[RCAST]{Research Center for Advanced Science and Technology, 
\\The University of Tokyo, 4-6-1, Komaba, Meguro-ku, Tokyo 153-8904, Japan}
\ead{feliciani@jamology.rcast.u-tokyo.ac.jp}

\address[Hongo]{Department of Aeronautics and Astronautics,
Graduate School of Engineering, The University of Tokyo,
7-3-1 Hongo, Bunkyo-ku, Tokyo 113-8656, Japan}

\author[RCAST,Hongo]{Daichi Yanagisawa}
\ead{tDaichi@mail.ecc.u-tokyo.ac.jp}

\author[RCAST,Hongo]{Katsuhiro Nishinari}
\ead{tknishi@mail.ecc.u-tokyo.ac.jp}

\begin{abstract}

In this paper, controlled experiments have been conducted to make deep analysis on the obstacle evading behavior of individual pedestrians affected by one obstacle. Results of Fourier Transform show that with the increase of obstacle width, the frequency and amplitude of body sway would barely be affected while the lateral deviation of walking direction would largely increase. On the one hand, the relation among the extracted gait features including body sway, stride length, frequency and speed has been illustrated. On the other hand, the walking direction can be featured by three critical evading points where apparent change of walking direction could be observed. Gaussian function has been used to fit the walking direction, thus allowing to estimate the three critical points and examine their variation with the increase of obstacle size. Furthermore, the direct-indirect evading and left-right turning preference as well as the possible psychological motivations behind have been analyzed. It is indicated that direct-evading pedestrians have a higher walking efficiency and right-turning pedestrians have a stronger tendency to behave `direct'. Results of this paper are expected to provide practical evidence for the modeling of pedestrian dynamics affected by obstacles.  

\end{abstract}


\end{frontmatter}


\section{Introduction}\label{sec:intro}


Pedestrians in real life always have to evade obstacles like walls, pillars and interior furnishings. Therefore, it is essential to understand the influence of obstacle on pedestrian dynamics and therefore evaluate or improve the design of obstacles to guarantee the comfortability of normal walking and the security of emergency evacuation.  

So far, many scholars have conducted experimental and simulation research to explore the influence of obstacle on crowd dynamics. For instance, experimental results show that putting an obstacle in an intersection of pedestrians could improve the stability of self-organized lanes under different conditions \cite{intersectionped}. One interesting phenomenon is that in evacuation scenario where pedestrians are urgent to get out the exit, putting an obstacle in front of the exit could help improve the outflow efficiency under certain conditions \cite{daichi,daichi2,zhaoobs,GAfrank,review}. Other reasonable obstacle settings that could help improve the evacuation efficiency have also been evaluated based on simulation \cite{obssimulation,jiangobs}. This phenomenon tends to be more apparent in the case of silo \cite{silo}, sheep \cite{sheep} and mice \cite{mice} outflow without self-awareness or with high competition, while hard to be observed in non-emergency egress conditions \cite{obsdecrease}. 

However, the extreme focus on crowd features leads to the deficiency in exploring the individual characteristics or motivations of the evading behavior. In particular, the evading rules in most present agent-based simulation models are unverified. 
Many classical models such as the cellular automata model \cite{ca,cacollision}, social force model \cite{Helbing2000} and velocity-based model \cite{vomodel1,vomodel2} have been modified to enable the obstacle-evading behavior. For instance, agents in some research were defaulted to choose the shortest route to the exit and their desired directions would directly point to the edge of the bottleneck nearest to the exit \cite{GAfrank,zhaoobs,guo,zhukongjin}. Some research reproduced the evading process through introducing a force lateral to the obstacle \cite{rudloff,guo,mine}. Other referential ways to avoid general collisions include using the Voronoi diagram to calculate the probability for the agent to detour to different directions \cite{sfmvoronoi}, describing the collision avoidance in the route choice process using game theory \cite{asano} and applying self-stopping mechanism to prevent collision \cite{parisi}. 

One essential problem when setting rules is the timing that a pedestrian begins to evade the obstacle. The evading behavior will be triggered at the beginning if agents choose to walk at the shortest distance \cite{GAfrank,zhaoobs,guo}. The concept of triggering condition is blurred in force-based models because the forces, though always exist, would be too small to affect the agents that are far away from the obstacle \cite{rudloff}. Although some studies have provided the triggering conditions, e.g. critical evading distance \cite{mingtang} or expected collision time \cite{mine}, their experimental evidence were still insufficient. 

Since simple rules of single agent could largely affect the crowd phenomenon \cite{Moussaid2}, it is of significance to provide reliable experimental analysis on the evading behavior in order to reproduce the influence of obstacle on crowd dynamics. However, to the author's knowledge, few experimental validation of evading rules using real obstacle have been proposed. Considering that the obstacle evading behavior is one particular case of collision avoidance behavior, we would like to name some experimental results on pairwise avoidance of pedestrians.

Some studies considered a standing pedestrian as obstacle and reported the critical headway, i.e. the distance between a pedestrian to the obstacle when he would like to begin evading the obstacle. The value of critical headway tended to be clustered between 0.9 m and 2.0 m in Ref. \cite{Lv} and between 1.5 m and 2.5 m in Ref. \cite{Moussaid}. The minimum avoidance distance was reported to be no larger than 1 m in Ref. \cite{trajfitting}. Besides, it has been revealed that the average critical walking space was about 2 m$^2$ for avoiding a standing person and 2.64 m$^2$ for avoiding an opposite moving pedestrian \cite{walkingspace}. In other conditions when avoiding a moving pedestrian, the variation of crossing angle would also influence the evading distance \cite{trajfitting,angleinteraction}. However, accurate methods to judge the exact location that a pedestrian begins to evade is still insufficient. 

Analysis about evading trajectories have also been conducted. In real walking process, pedestrian trajectories were reported to contain two kinds of information: the body sway and the main walking direction \cite{trajfitting,gaitcircle}.  The body sway amplitude of pedestrian crowd heading to the bottleneck tended to be about 0.04 to 0.06 m and showed a negative correlation with the velocity which ranged from 0.5 m/s to 1.5 m/s \cite{Hoogendoorn2005}. To almost free pedestrians, the amplitude tended to be 0.036$\pm$0.024 m and the sway period was 1.60$\pm$0.28 m \cite{trajfitting}. With regard to the evading direction, some studies used quadratic equation \cite{trajfitting} or modified force-based functions \cite{seer,pairwise} to fit the main walking trajectory, or directly obtained the trajectory through filtering the body sway \cite{gaitcircle}.




In the above research, few have studied the evading behavior of pedestrians when encountering real obstacle, while the shape and material of the obstacle may largely affect the pedestrian behavior. In particular, panel-like obstacles like walls are often used either as architectural boundaries or as obstacles to fulfill certain service functions, while their influence on individual pedestrians have not been noticed yet. 

Therefore, we would like to thoroughly analyze the evading behavior of single pedestrians when confronting with a panel-like obstacle. The obstacle size will be adjusted to see the corresponding influence, and a reliable method will be proposed to fit the evading trajectories and help judge the exact location that a pedestrian begins to evade. The gait features and evading preference of participants will also be analyzed.  
\section{Experiment setting}\label{sec:experiment}
\begin{figure}[ht]
\centering
\subfigure[experiment scenario]{
\label{f1:a} 
\includegraphics[width=0.45\textwidth]{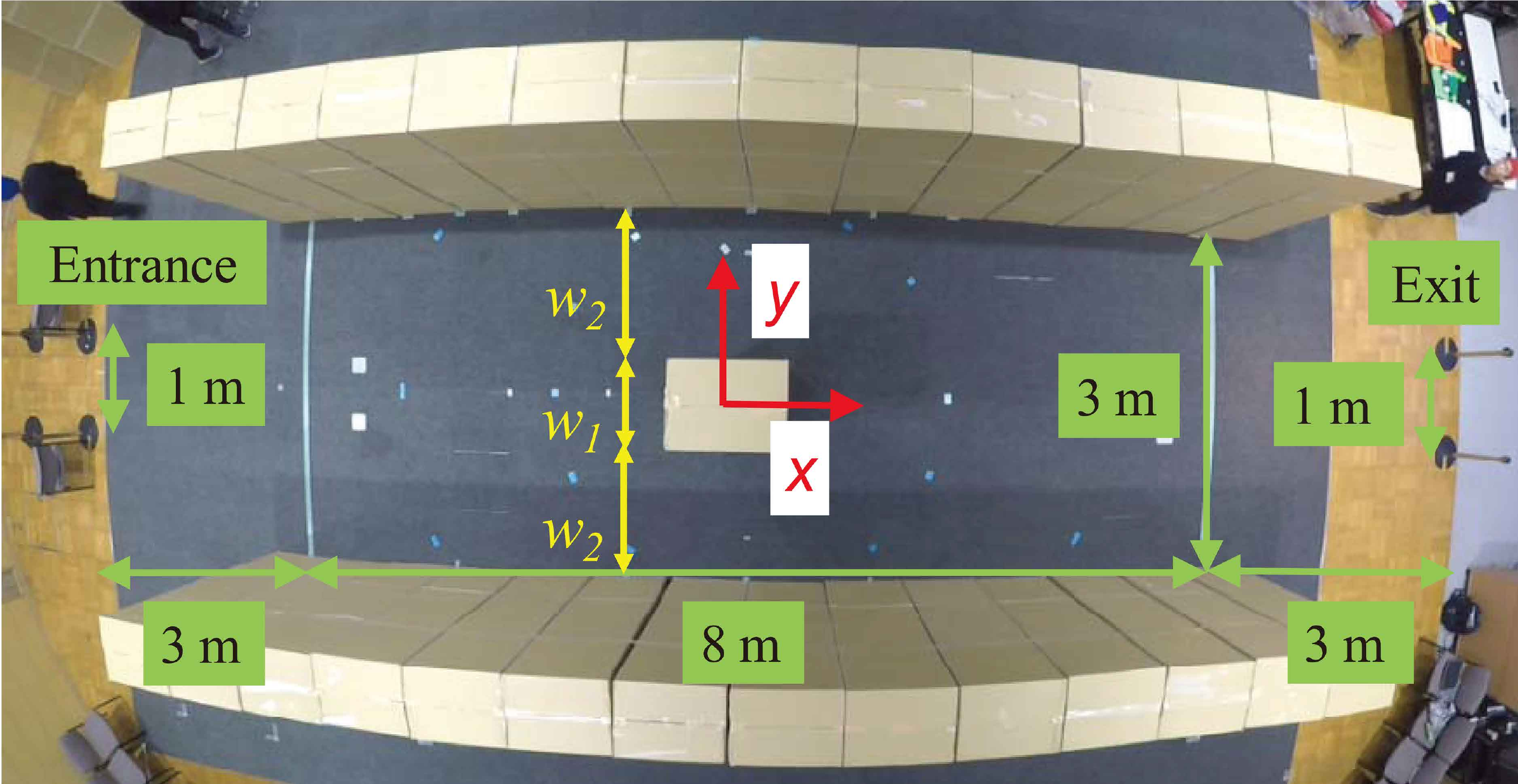}}
\subfigure[geometrical setting]{
\label{f1:b} 
\includegraphics[width=0.45\textwidth]{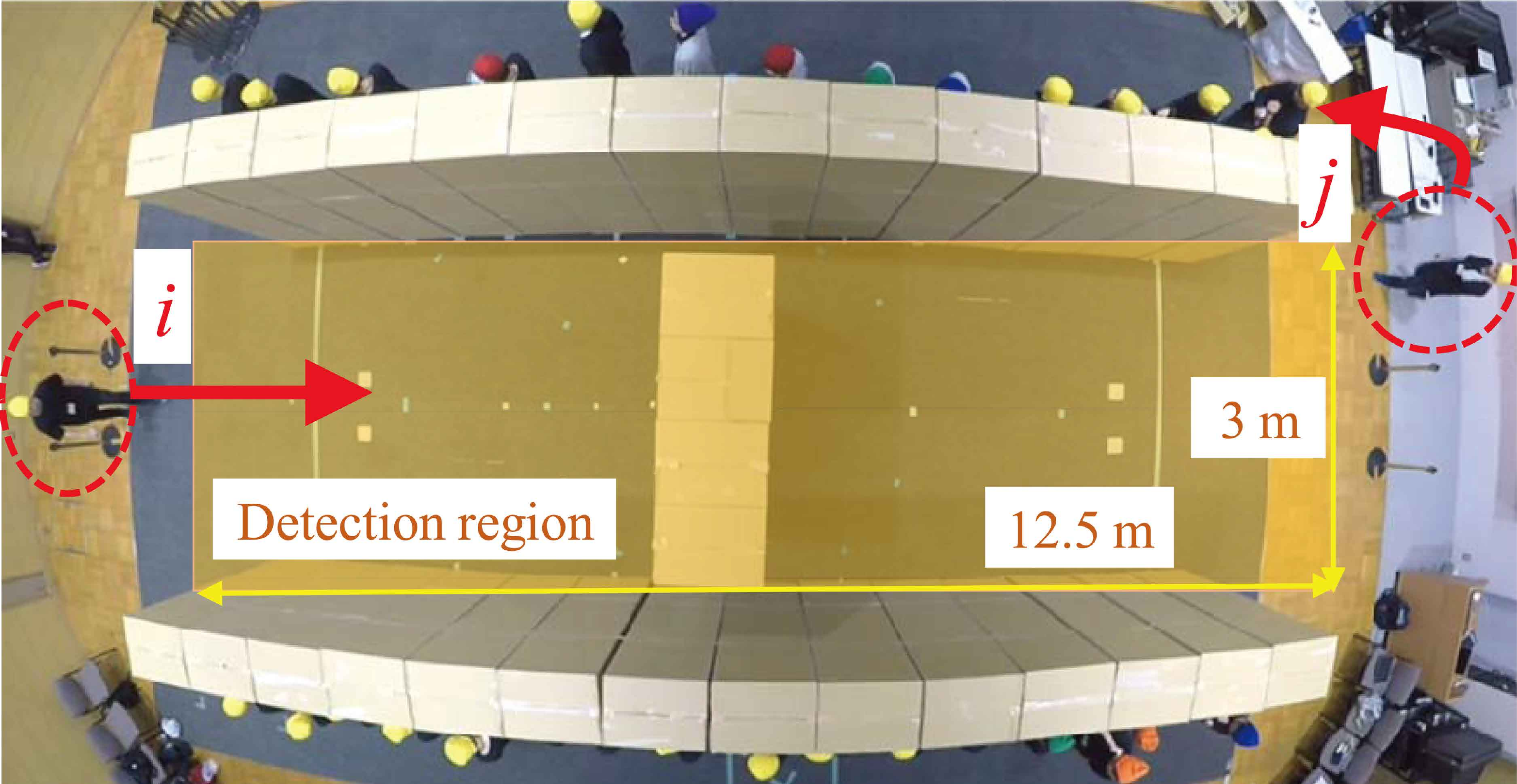}}
\caption{Experimental scenario: a corridor with a wall-like obstacle in the middle.}
\label{f1} 
\end{figure}

Experiments have been conducted to explore the influence of obstacle on the behavior of single pedestrians. The experiment scenario was set as a corridor with a wall-like obstacle whose width would be adjusted. The experiments was conducted on November 4th, 2017 in the Lecture hall of RCAST Building 4, The University of Tokyo, Japan. In total, 32 male students whose ages ranged from 20 to 25 have participated in the experiments. As is shown in Fig. \ref{f1:a}, we used cardboard boxes to construct the walls of the corridor as well as the obstacle in the middle of the corridor. The depth, width and height of each cardboard box was 0.42, 0.62 and 2.19 m respectively. Other geometrical information including the coordinate system of the experiment scenario can also be seen in Fig. \ref{f1:a}. 

The obstacle formed two equal-width bottlenecks with the two walls of the corridor. We define the width of the obstacle as $w_1$ and the width of each bottleneck as $w_2$. The number of boxes that form the obstacle is defined as $box$, and the relations among $w_1$, $w_2$ and $box$ are shown in Eq. \ref{eq:e1}. 
\begin{equation}
\begin{split}
\label{eq:e1}
w_1=0.42\  {\rm m}*box,   \\
w_2=(3 \ {\rm m}-w_1)/2
\end{split}
\end{equation}

Due to the limitation of corridor width, there are mostly four boxes that compose the obstacle in our experiments. The values of $w_1$ and $w_2$ under each possible value of $box$ can be seen in Tab. \ref{tab:t1}. Essentially, if we increase the value of $box$, $w_1$ will increase and $w_2$ will decrease correspondingly. Therefore, we will use $box$ to show the variation of obstacle and bottleneck width hereinafter. 

\begin{table}[ht]
\centering
\caption{All the possible values of $box$, $w_1$ and $w_2$ in our experiments.}
\label{tab:t1}
\begin{tabular}{|c|c|c|c|c|c|}
\hline
   $box$ & 0 & 1 & 2 & 3 & 4 \\
   \hline
   $w_1 [m]$ & 0 & 0.42 & 0.84 & 1.26 & 1.68 \\
    \hline
    $w_2 [m]$ & 1.5 & 1.29 & 1.08 & 0.87 & 0.66 \\
    \hline
\end{tabular}
\end{table}

A camera was set above the horizontal axis of the corridor and fixed 6 meters above the ground. Recordings of the camera was adjusted to full HD mode (1920$\times$1080 pixel) with a frame rate of 30 fps. With the videos of the experiments as rough data, the recognition and tracking of pedestrians could be achieved using PeTrack software (version 0.8) \cite{petrack}. Pedestrians were required to wear colored caps so that their positions at each video frame could be detected. Meanwhile, pedestrians would wear black or grey shirts to avoid error detection. 

Five sub-experiments on the evading behavior of single pedestrians have been conducted with $box$ being 0 (no obstacle), 1, 2, 3 and 4 respectively. As is shown in Fig. \ref{f1:b}, the 32 pedestrians were evenly distributed behind the two walls and required to stand in two queues before the experiment. 

During the experiment, pedestrians were asked to pass by the corridor from the entrance to the exit one by one. To be specific, a pedestrian was only allowed to begin to walk after his former pedestrian passed by the corridor. As is shown in Fig. \ref{f1:b}, pedestrian $i$ was instructed to pass the entrance after his former pedestrian $j$ has passed the exit, and pedestrian $j$ would go back to his former queue to keep clear of the exit. Please note the obstacles and walls were taller than the pedestrians to prevent pedestrians from observing or imitating the behavior of others. Locations of pedestrians within the detection region (about 12.5$\times$3 m$^2$) shown in Fig. \ref{f1:b} have been extracted, and finally 32$\times$5=108 sets of location data could be obtained.
\section{Gait features of evading pedestrians}\label{sec:common}
\begin{figure}[ht]
\centering
\subfigure[$box$=0]{
\label{traj:a} 
\includegraphics[width=0.45\textwidth]{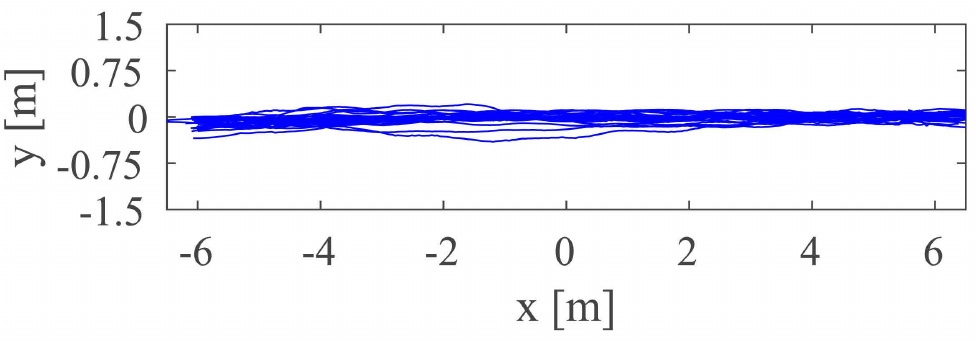}}
\subfigure[$box$=1]{
\label{traj:b} 
\includegraphics[width=0.45\textwidth]{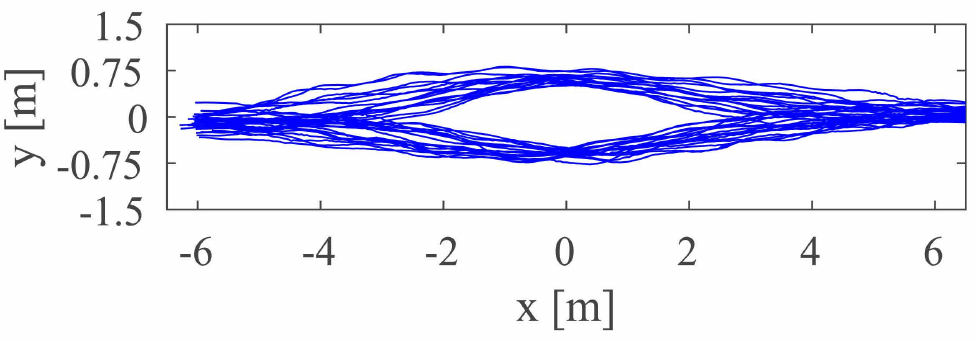}}
\hspace{1in} 
\subfigure[$box$=2]{
\label{traj:c} 
\includegraphics[width=0.45\textwidth]{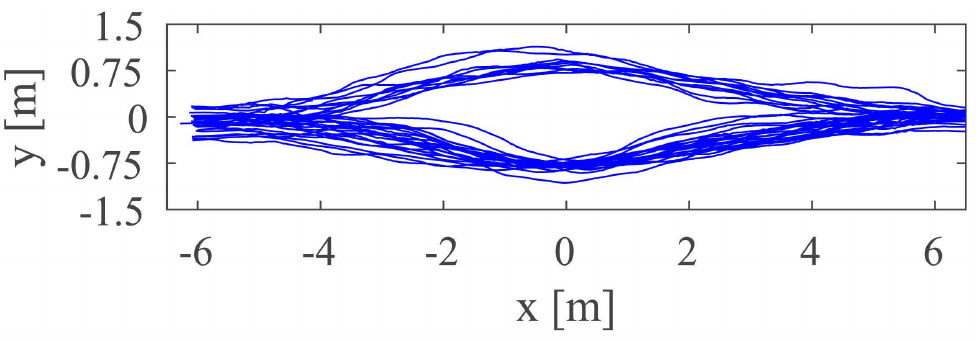}}
\subfigure[$box$=3]{
\label{traj:d} 
\includegraphics[width=0.45\textwidth]{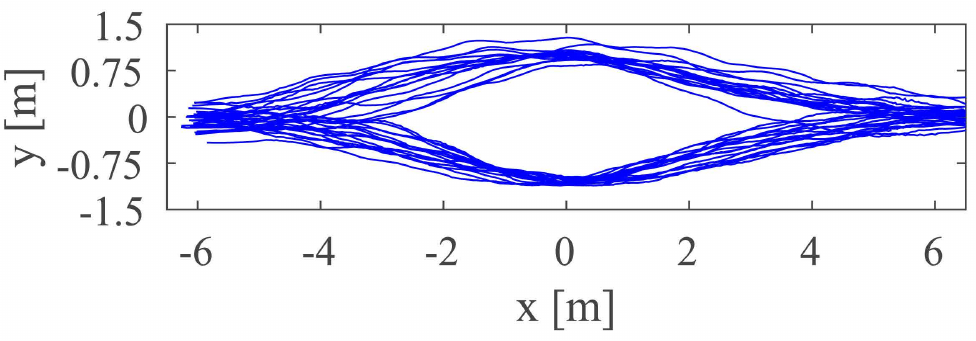}}
\hspace{1in} 
\subfigure[$box$=4]{
\label{traj:e} 
\includegraphics[width=0.45\textwidth]{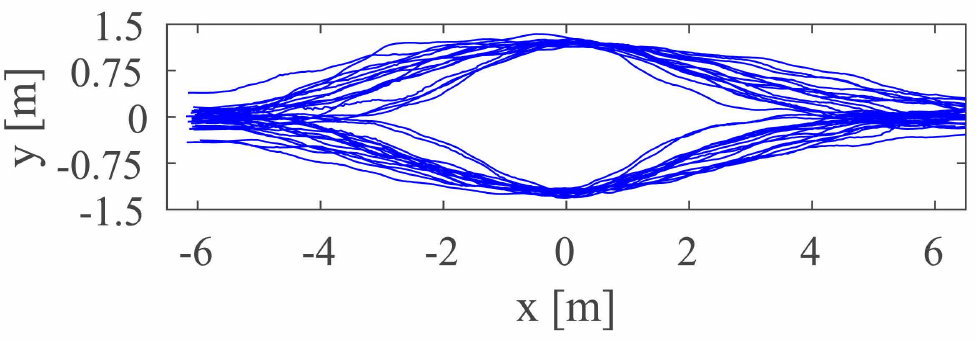}}
\caption{All the trajectories of 32 pedestrians under five sub-experiments with different obstacle width.}
\label{traj} 
\end{figure}

For a certain pedestrian, the variation of $y$ coordinate with the $x$ coordinate could be obtained and his trajectory could be depicted through connecting all the neighboring location points. For general impression, we plot the trajectories of all the 32 pedestrians under five kinds of obstacle sizes as shown in Fig. \ref{traj}. 

When there is no obstacle in the corridor, pedestrians tend to walk horizontally along the $x$ axis as shown in Fig. \ref{traj:a}. When an obstacle exists in the corridor, pedestrians have to deviate the walking directions from the $x$ axis to avoid collision as shown in Fig. \ref{traj:b}-\ref{traj:e}. To further analyze the variation trend of pedestrian trajectory, the slope of the trajectory curve would be calculated. We assume that there are $n$ video frames in one test, and the value of slope of pedestrian $i$ at a certain frame $t$ could be calculated based on his coordinates at the five neighboring frames. The slope of pedestrian $i$ at frame $t$ is defined as $k_i (t)$, which can be calculated by Eq. \ref{eq:slo}. 

\begin{equation}
\label{eq:slo}
k_i (t)=\frac{y_i (t+2)-y_i (t-2)}{x_i (t+2)-x_i (t-2) }, \quad  (3\leq t\leq n-2,x_i (t+2)\neq x_i (t-2))   
\end{equation}

\begin{figure*}[ht]
\centering
\subfigure[$box$=0]{
\label{slope:a} 
\includegraphics[width=0.45\textwidth]{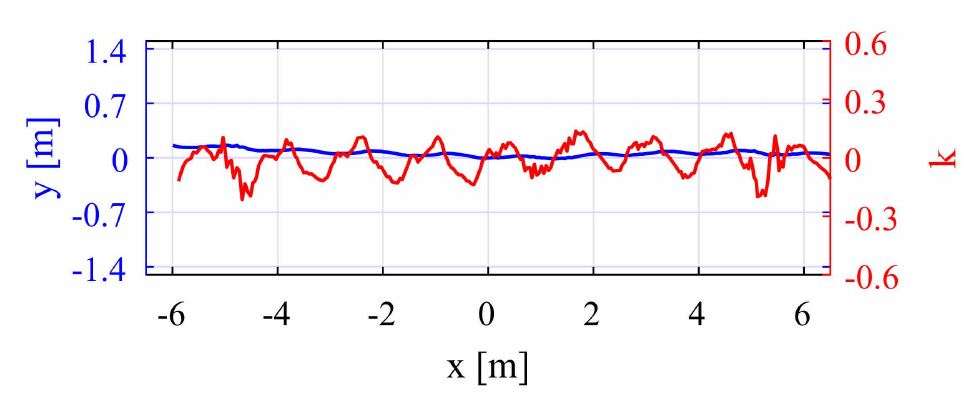}}
\subfigure[$box$=1]{
\label{slope:b} 
\includegraphics[width=0.45\textwidth]{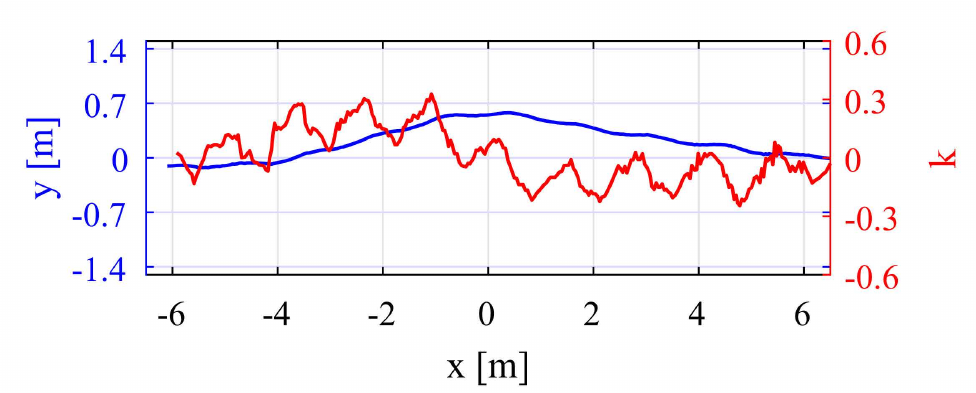}}
\hspace{1in} 
\subfigure[$box$=2]{
\label{slope:c} 
\includegraphics[width=0.45\textwidth]{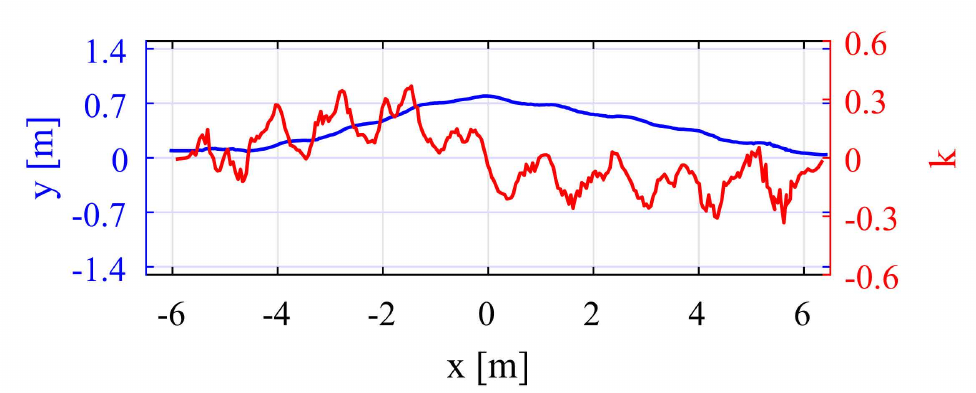}}
\subfigure[$box$=3]{
\label{slope:d} 
\includegraphics[width=0.45\textwidth]{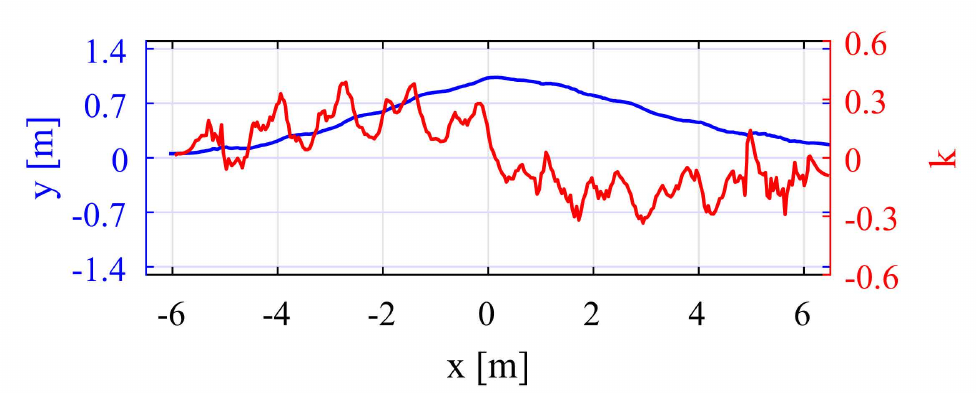}}
\hspace{1in} 
\subfigure[$box$=4]{
\label{slope:e} 
\includegraphics[width=0.45\textwidth]{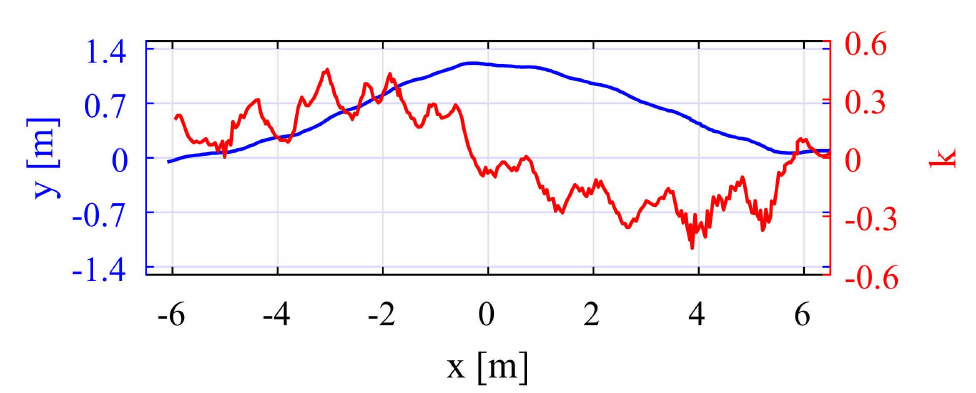}}
\caption{The variation of $y$ coordinate and slope against $x$ coordinate of a certain participant.}
\label{slope} 
\end{figure*}

According to Eq. \ref{eq:slo}, the variation of slope value $k$ during the walking process of each pedestrian can be calculated. Through comparing the trajectory and slope data of all the pedestrians, certain behavior pattern can be observed. Taking a typical participant as an example, the variation of $y$ coordinate and slope $k$ against $x$ coordinate under different obstacle sizes could be seen in Fig. \ref{slope}. 

\begin{figure}[ht]
\centering
\subfigure[$box$=0]{
\label{fourier:a} 
\includegraphics[width=0.3\textwidth]{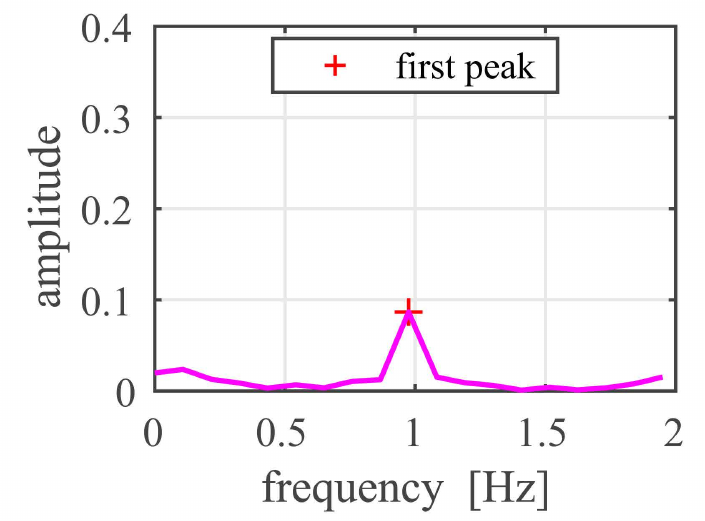}}
\subfigure[$box$=1]{
\label{fourier:b} 
\includegraphics[width=0.3\textwidth]{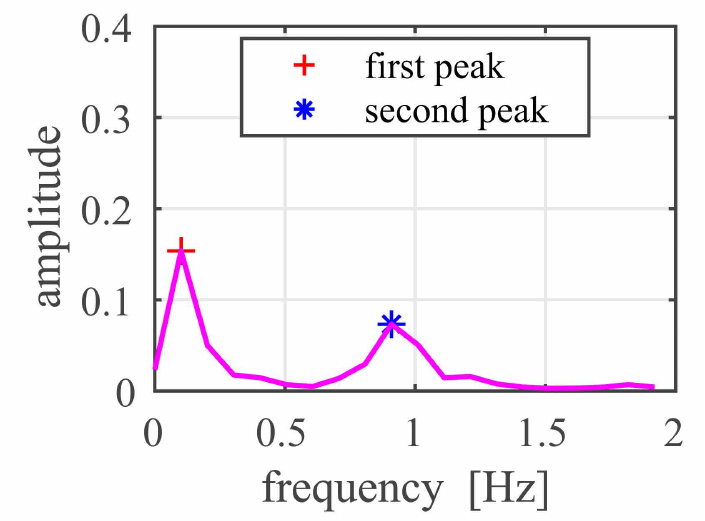}}
\subfigure[$box$=2]{
\label{fourier:c} 
\includegraphics[width=0.3\textwidth]{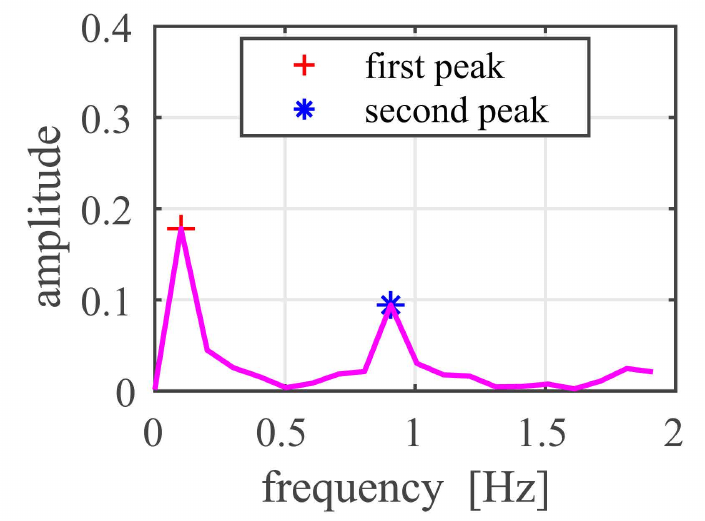}}
\hspace{1in} 
\subfigure[$box$=3]{
\label{fourier:d} 
\includegraphics[width=0.3\textwidth]{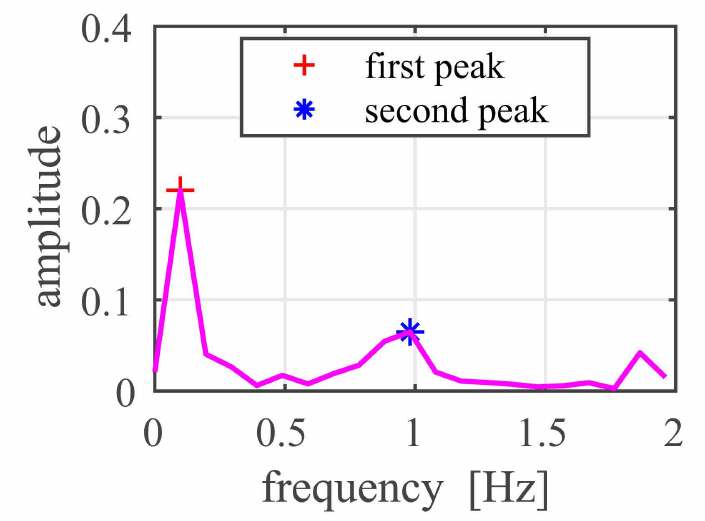}}
\subfigure[$box$=4]{
\label{fourier:e} 
\includegraphics[width=0.3\textwidth]{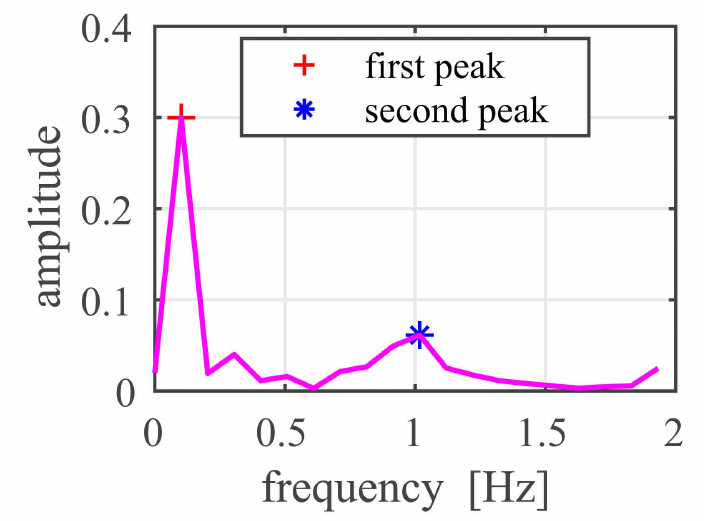}}
\caption{Results of Fourier Transform based on the slope curve.}
\label{fourier} 
\end{figure}

When there is no obstacle (see Fig. \ref{slope:a}), a pedestrian will walk straight and the corresponding slope curve behaves like periodic wave. We presume the periodic wave is caused by the natural oscillation of human beings without disability who will step out his left and right legs in shift to move forward. When there is an obstacle (see Fig. \ref{slope:b}-\ref{slope:e}), a pedestrian has to change his walking direction, making the trajectory a unimodal curve. Nevertheless, his body will also oscillate and a seemingly periodic wave can also be observed in the slope curve. For a better illustration, we use Fourier Transform to obtain the frequency and amplitude characteristics of the slope curve and the results can be seen in Fig. \ref{fourier}. 

It turns out that two peaks can be observed when $box \geq 1$ (see Fig. \ref{fourier:b}-\ref{fourier:e}), which means the slope curve can be roughly fitted by two trigonometric functions. The first peak can be regarded as the amplitude for the slope curve of walking direction, and the amplitude becomes larger with the increase of obstacle size because the pedestrian has to detour a longer lateral distance. The second peak can be regarded as the amplitude for the slope wave of body sway, and its frequency keeps at about 0.9 Hz and the corresponding amplitude keeps around 0.08 m despite the variation of obstacle width. The peak representing body sway can also be observed in no-obstacle condition as first peak (see Fig. \ref{fourier:a}). Therefore, we could presume that the body sway of an undisturbed pedestrian would not change too much with the variation of obstacle width, which can be validated through analyzing the variation of gait features. 


The gait features of pedestrians that would be analyzed include sway amplitude, stride frequency, stride length and walking speed during the evading process. In some present research \cite{parisi} \cite{Hoogendoorn2005}, the body sway amplitude and stride length are usually calculated based on the gap between real trajectory and fitted trajectory. Other research \cite{gaitcircle} use radial coordinate other than trajectory curve to obtain the gait features, which is suitable for pedestrians walking in a circle so that the walking direction can be decoupled. In this study, we would like to try a different way to estimate the gait features based on the slope curves rather than the trajectory curves. In our method, the body sway are no longer sensitive to the function used to fit the walking direction but completely depends on its own features. Besides, the body sway is more apparent in slope curve than trajectory curves (See Fig. \ref{slope}). 

\begin{figure}[ht]
\centering
\subfigure{
\includegraphics[width=0.45\textwidth]{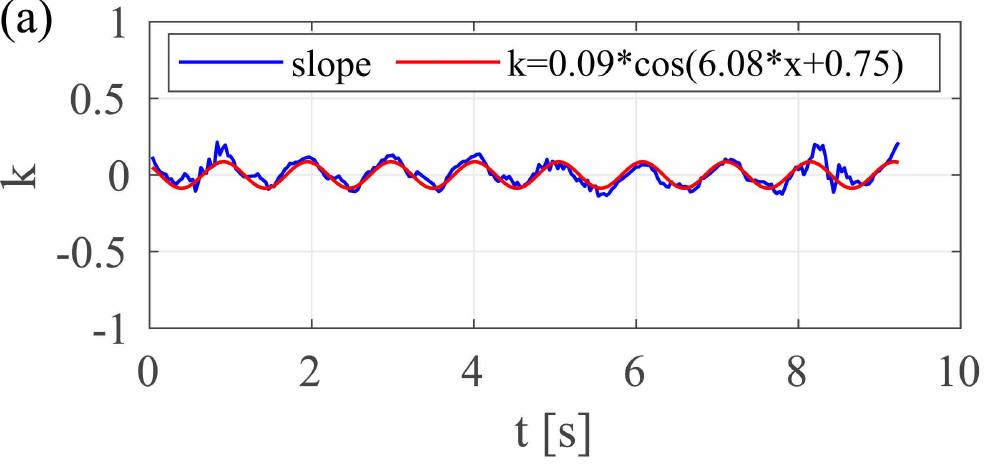}}
\subfigure{
\includegraphics[width=0.45\textwidth]{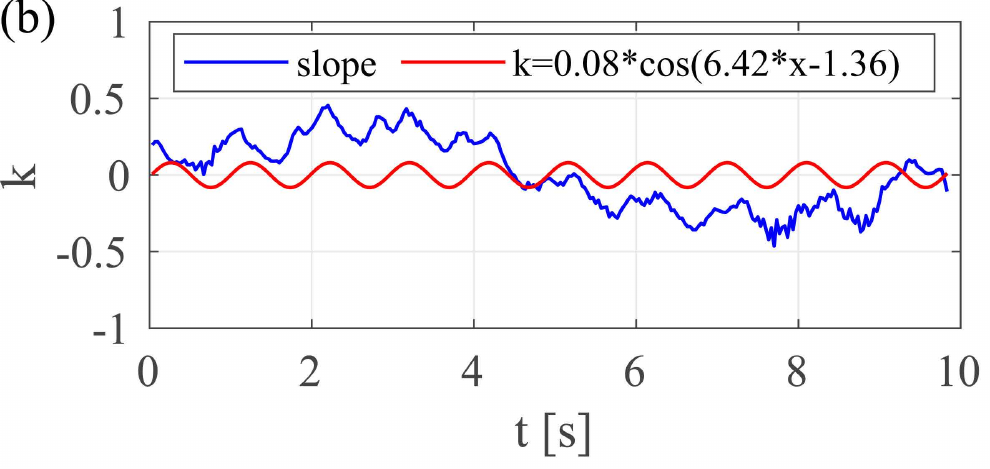}}
\hspace{1in} 
\subfigure{
\includegraphics[width=0.45\textwidth]{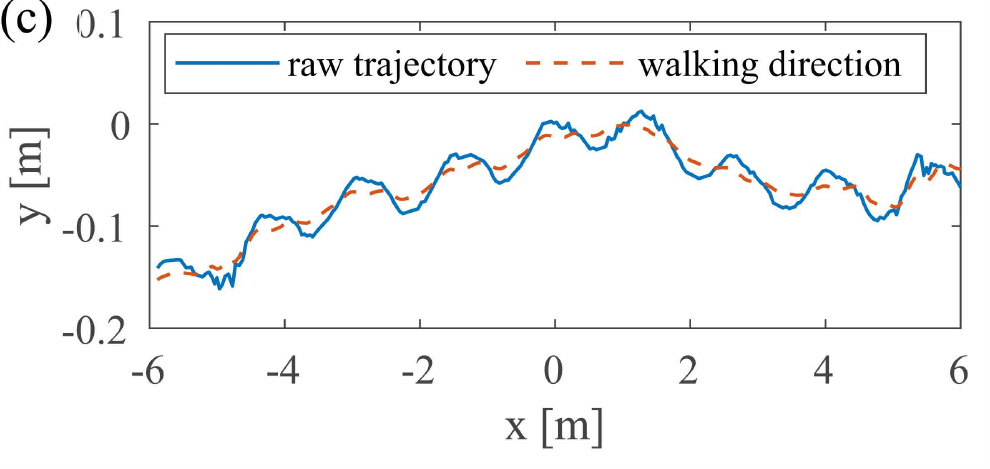}}
\subfigure{
\includegraphics[width=0.45\textwidth]{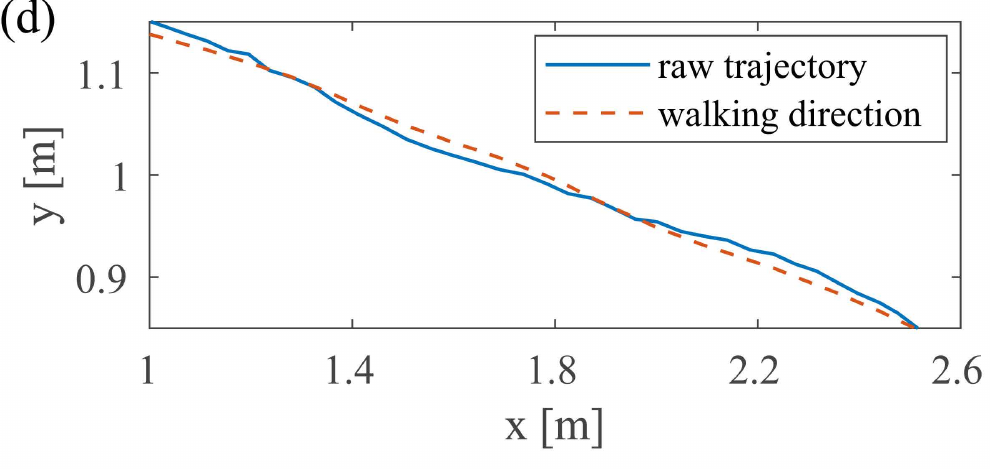}}
\caption{Fitting the oscillation of slope and trajectory curve caused by body sway. (a) and (c) are respectively the slope and trajectory curve when $box=0$, while (b) and (d) correspond to $box=4$. Please note that the range of $x$ coordinate in (d) is set much smaller than (c) to make the body sway more visible considering the lateral deviation of walking direction curve is much larger than that of body oscillation when $box=4$.}
\label{oscillate} 
\end{figure}

The amplitude and frequency of the slope curve induced by body sway could be extracted through Fourier Transform results. Therefore, we would use a cosine function with the extracted amplitude and frequency to fit the oscillation of slope curve. Examples of raw and fitted slope curve of body sway can be seen in Fig. \ref{oscillate} (a) and \ref{oscillate}(b). The oscillation of trajectory curve can therefore be obtained through calculating the integral of slope curve, and the walking direction could be obtained through subtracting the oscillation from the trajectory curve. The outcome can be seen in Fig. \ref{oscillate}(c) and \ref{oscillate}(d). 


Based on the method above, we could obtain the amplitude and stride frequency of all the pedestrians. For each cycle of body sway, a pedestrian will walk for two steps. We then assume the distance that a pedestrian step out for two steps as stride length and calculate the walking speed as the product of stride frequency and stride length. The calculation results can be seen in Tab. \ref{tab:gait}.
\begin{table}
  \centering
  \caption{The gait information of individual pedestrians under different obstacle sizes.}
    \begin{tabular}{|c|c|c|c|c|c|c|c|c|}
    \hline
    \multirow{2}[3]{*}{\textit{box}} & \multicolumn{2}{c|}{amplitude} & \multicolumn{2}{c|}{stride frequency} & \multicolumn{2}{c|}{stride length} & \multicolumn{2}{c|}{speed} \\
\cline{2-9}          
& \multicolumn{1}{p{4em}<{\centering}|}{mean\newline{}value[m]} & \multicolumn{1}{p{4em}<{\centering}|}{standard\newline{}deviation} & \multicolumn{1}{p{4em}<{\centering}|}{mean\newline{}value[Hz]} & \multicolumn{1}{p{4em}<{\centering}|}{standard\newline{}deviation} & \multicolumn{1}{p{4em}<{\centering}|}{mean\newline{}value[m]} & \multicolumn{1}{p{4em}<{\centering}|}{standard\newline{}deviation} & \multicolumn{1}{p{4em}<{\centering}|}{mean\newline{}value[m/s]} & \multicolumn{1}{p{4em}<{\centering}|}{standard\newline{}deviation} \\
    \hline
    0     & 0.0149 & 0.0146 & 0.979 & 0.076 & 1.484 & 0.264 & 1.426 & 0.143 \\
    \hline
    1     & 0.0151 & 0.0054 & 0.947 & 0.092 & 1.546 & 0.326 & 1.442 & 0.164 \\
    \hline
    2     & 0.0150 & 0.0050 & 0.960 & 0.070 & 1.487 & 0.149 & 1.425 & 0.161 \\
    \hline
    3     & 0.0150 & 0.0048 & 0.960 & 0.073 & 1.458 & 0.164 & 1.394 & 0.140 \\
    \hline
    4     & 0.0156 & 0.0062 & 0.941 & 0.081 & 1.471 & 0.155 & 1.380 & 0.154 \\
    \hline
    \end{tabular}%
  \label{tab:gait}%
\end{table}%

Despite the four types of mean values in Tab. \ref{tab:gait} would fluctuate with the increase of $box$, we assume the fluctuation is allowable in one standard deviation error range and hence validate that the gait features would not change too much with the increase of obstacle width. One possible reason is that pedestrians all walk at their desired pace and would not suddenly accelerate or decelerate because of sudden change of walking environment. On the other hand, if pedestrians are under crowd conditions where pedestrians have to handle with complicated mutual interactions, the gait information may be difficult to be extracted.

\begin{figure}
\centering
\subfigure[stride length-speed relation  ($R^2=0.562$)]{
\label{stride:a} 
\includegraphics[width=0.45\textwidth]{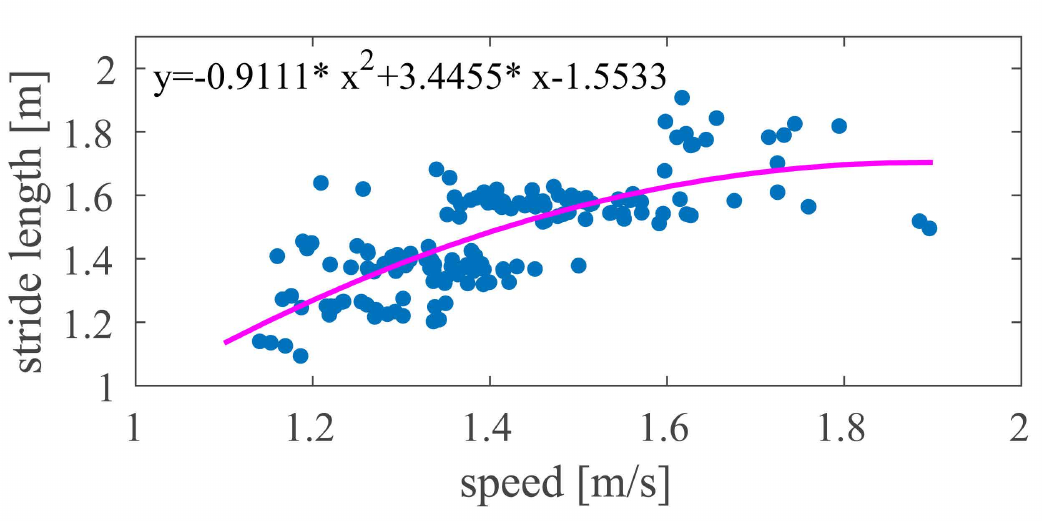}}
\subfigure[frequency-speed relation ($R^2=0.152$)]{
\label{stride:b} 
\includegraphics[width=0.45\textwidth]{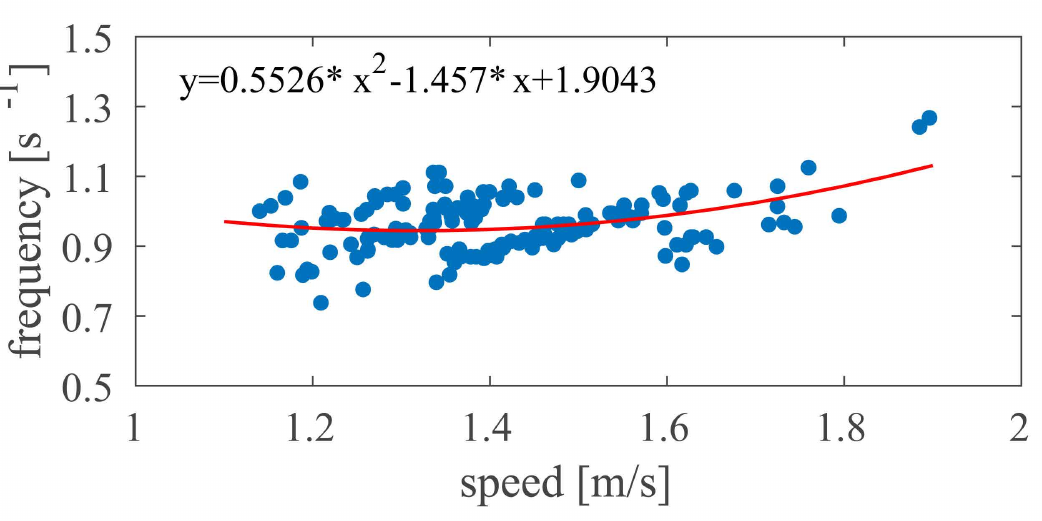}}
\hspace{1in} 
\subfigure[stride length-frequency relation ($R^2=0.224$)]{
\label{stride:c} 
\includegraphics[width=0.45\textwidth]{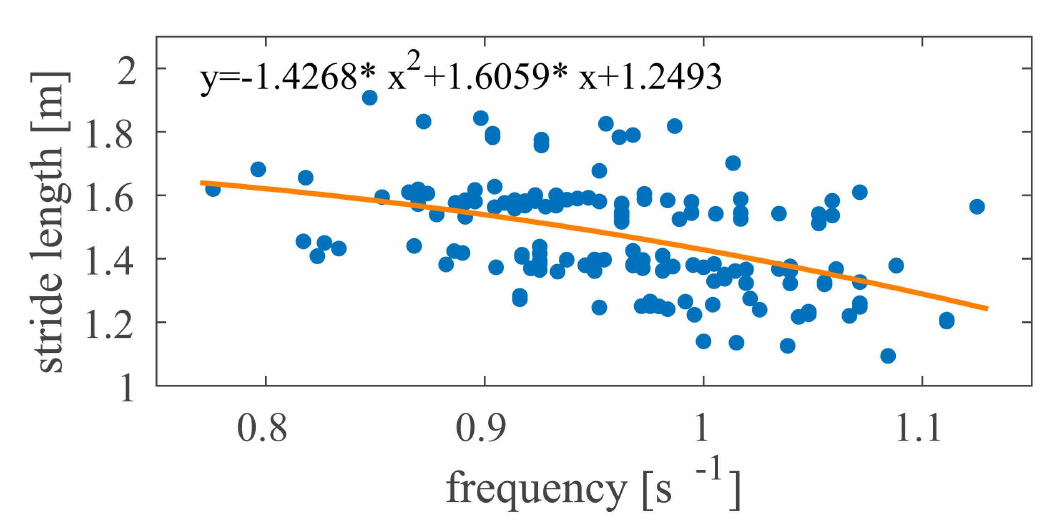}}
\subfigure[amplitude-frequency relation ($R^2=0.207$)]{
\label{stride:d} 
\includegraphics[width=0.45\textwidth]{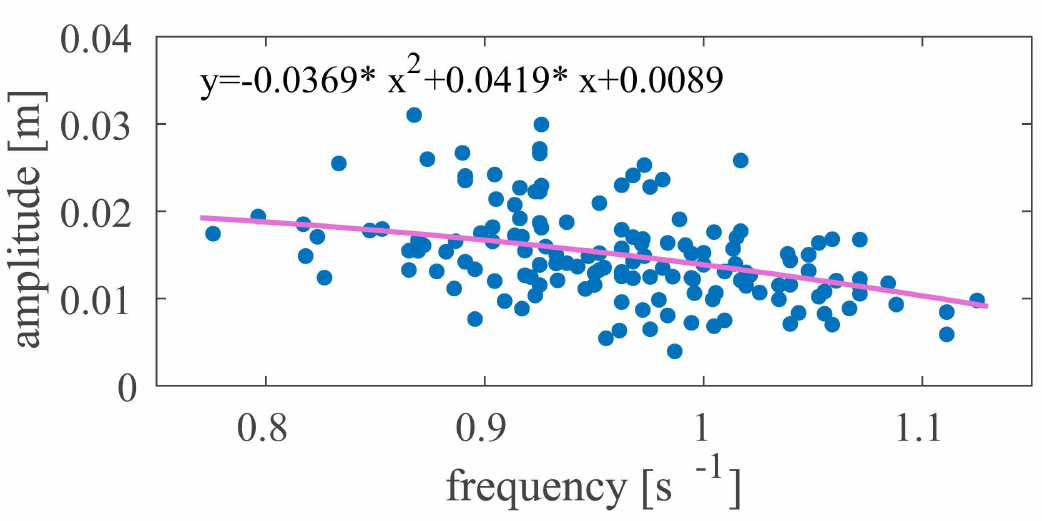}}
\caption{Relations among the gait features. The values of $R^2$ in the brackets represent the coefficient of determination for fitting.}
\label{stride} 
\end{figure}

After obtaining the three basic gait features, i.e., stride frequency, stride length and speed of the pedestrians, we would like to check their mutual correlations through calculating the Pearson correlation coefficient $c$. Results show that the speed is strongly related to stride length ($c=0.73$) and weakly related to stride frequency ($c=0.32$). It is hence indicated that pedestrians with high walking speed tend to have higher stride frequency and longer stride length. Compared with higher stride frequency, the increase of speed mainly attributes to longer stride length considering the correlation coefficient. Besides, the stride frequency and stride length have weak negative correlation ($c=-0.41$), which means the stride length will be smaller when the stride frequency becomes higher, or vise versa. We have used quadratic functions to describe the frequency-speed, stride length-speed and frequency-stride length relationship as shown in Fig. \ref{stride} (a-c). 

Moreover, the relation between the amplitude of body sway and the three basic gait features can be examined. Results show that the sway amplitude is weakly negatively correlated with stride frequency ($c=-0.45$), which means pedestrians with low stride frequency tend to have larger amplitude of body sway, or vise versa. The amplitude-frequency relation have been illustrated in Fig. \ref{stride:d}. Nevertheless, the amplitude of body sway is not or very weakly related with stride length ($c=-0.13$) and walking speed  ($c=0.19$).


\section{Fitting of trajectories and estimation of critical points}\label{sec:gaussian}
In this section, we would like to analyze the variation of walking direction during the evading process. We presume that the evading behavior can be featured by three critical points where apparent deviation of walking direction could be observed. The three points would be estimated through fitting the evading trajectory. 

\subsection{Critical evading points: MP, SP and EP}\label{sec:gaussian:walking_trajectory}

\begin{figure}[ht]
\centering
\includegraphics[width=0.6\textwidth]{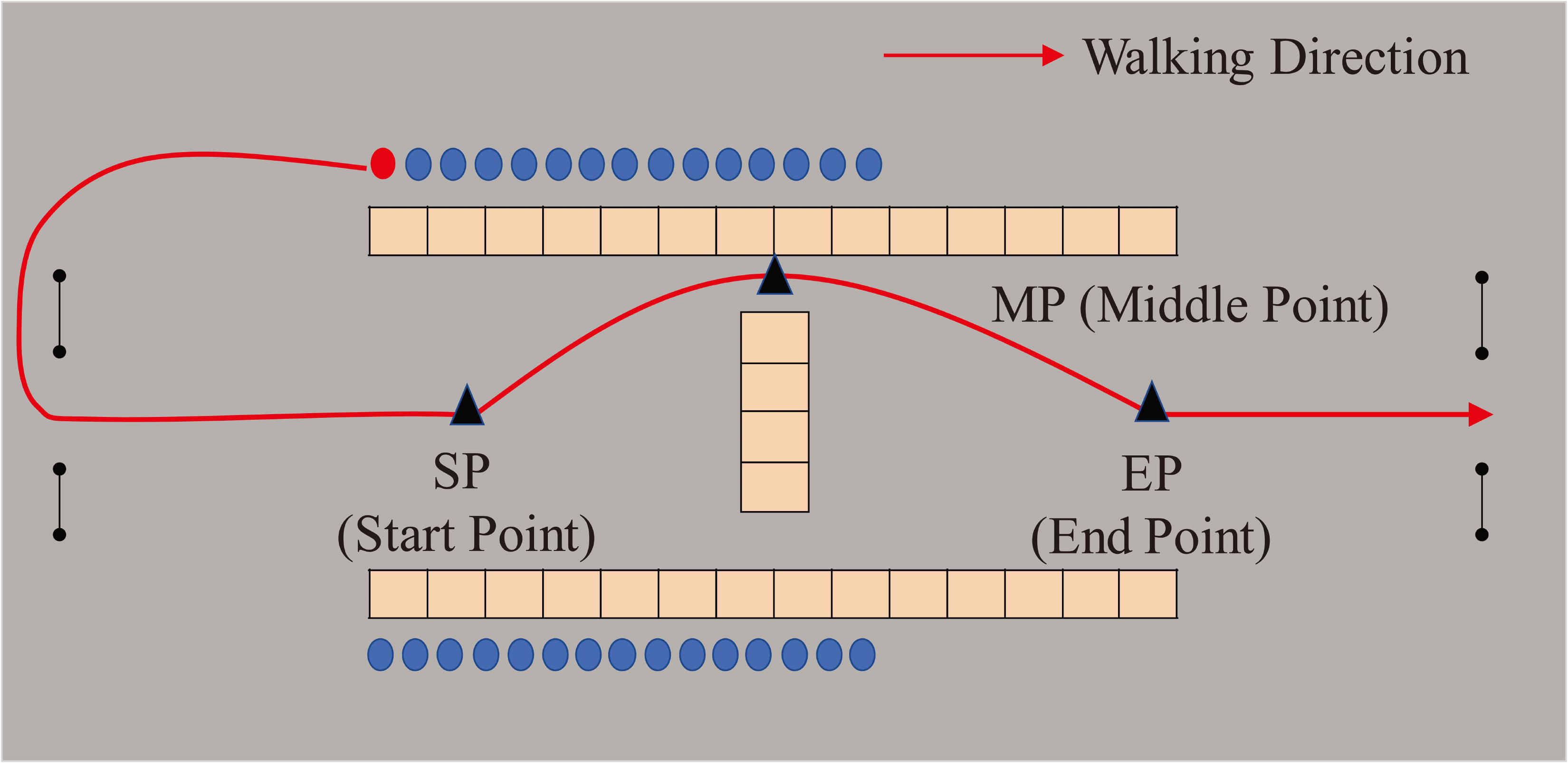}
\caption{Assumed walking trajectory and the three critical evading points.}
\label{spmpep}
\end{figure}

Based on the observation of the evading process and trajectories of single pedestrians, we assume that a pedestrian tends to evade an obstacle following the pattern shown in Fig. \ref{spmpep}. Generally, a pedestrian would change his rough walking direction at three critical locations that we define as Start Point (SP), Middle Point (MP) and End Point (EP) respectively. To be specific, when stepping into the corridor, a pedestrian tends to first walk straight and then start to evade the obstacle by changing his walking direction at SP. After passing by the obstacle successfully, he begins to change his walking direction again in order to traverse the corridor at MP. When heading to the exit, a pedestrian tends to first go back to the middle horizontal axis and then begin to walk straight towards the exit at EP. Please note that some pedestrians will directly change the walking direction when passing by the entrance or directly head to the exit after passing the MP. In this condition, the location of SP or EP is at the entrance or exit respectively. 

Based on the location of SP and EP in the experimental region, we assume there are roughly two types of behavior, i.e. `direct' and `indirect'. When getting into the corridor from the entrance, some pedestrians would directly change the walking direction to evade the obstacle, and we define them as `direct' evading pedestrians. On the other hand, some pedestrians would first walk straight for some distance and then change the walking direction, thus we define them as `indirect' evading pedestrians. 

Analogously, the two kinds of behavior can also be observed when pedestrians head to the exit after passing by the obstacle. Some pedestrians will directly head to the exit, and we name them as `direct' returning pedestrians. Other pedestrians will first return to the middle horizontal axis and then go straight until passing by the exit, and we name them as `indirect' returning pedestrians. Please note that the behavior of a certain pedestrian may not be the same when he is before and after the obstacle. For instance, `direct' evading pedestrians are not necessarily `direct' returning pedestrians.


\subsection{Estimation of MP}\label{sec:gaussian:mp}
\begin{figure}[ht]
\centering
\subfigure[distribution of $x$ coordinates]{
\label{mp:a} 
\includegraphics[width=0.45\textwidth]{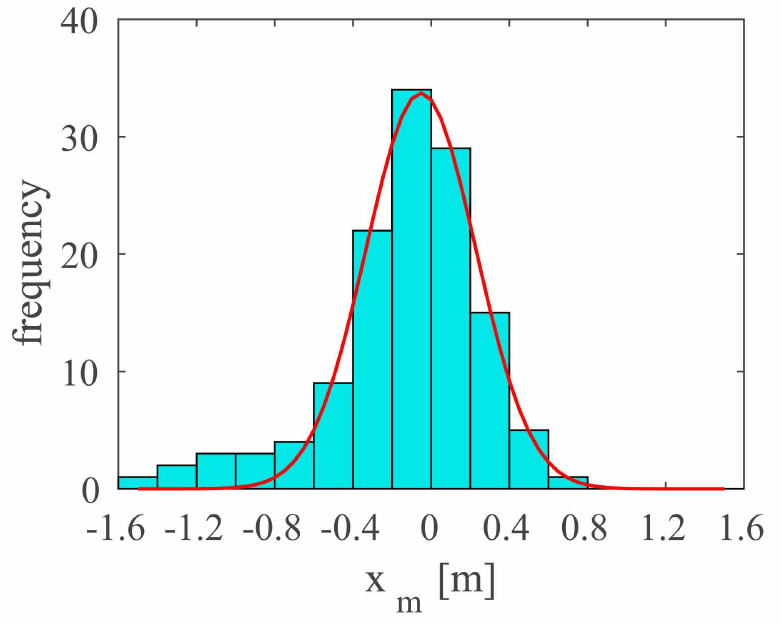}}
\subfigure[variation of $x$ coordinates]{
\label{mp:b} 
\includegraphics[width=0.45\textwidth]{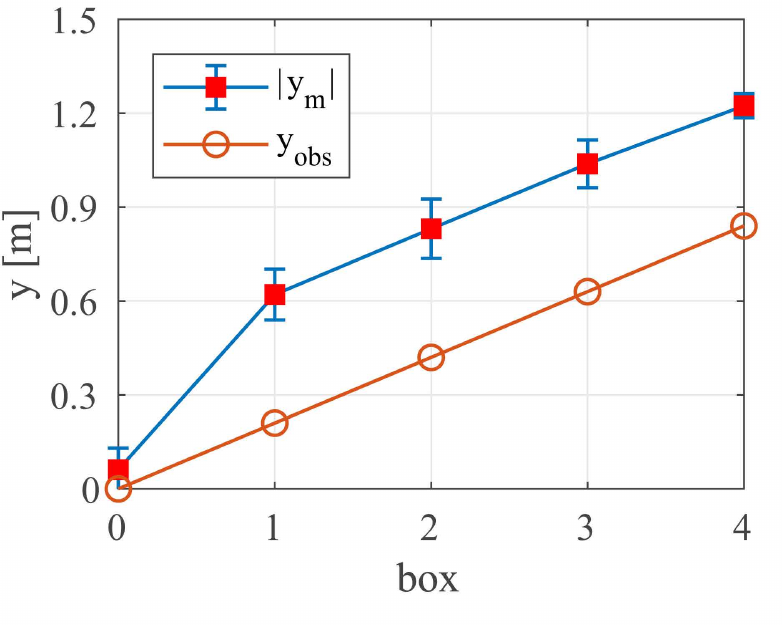}}
\caption{Features of the $x$ and $y$ coordinates of MP.}
\label{mp} 
\end{figure}

Considering the definition of MP, we regard it as the location with the largest deviation from the $y=0$ axis. For comparison, we assume the coordinate of MP as $[{x_{\rm m}},{y_{\rm m}}]$ and half of the obstacle width, i.e., the $y$ coordinate of the upper obstacle edge, as $y_{\rm obs}$. The features of the location coordinates of MP can be seen in Fig. \ref{mp}. Please note that $x_{\rm m}$ is not affected by the variation of obstacle width.

The value of $x_{\rm m}$ follows a normal distribution with the mean value being -0.052 m and the one standard deviation being 0.281 m as shown in Fig. \ref{mp:a}. The results of k-s test also proves the normality (p-value$=0.026<0.05$). On the other hand, the average value of $|y_{\rm m}|$ under each obstacle width and the corresponding error bar representing one standard deviation range could be obtained. For comparison, we also depict the corresponding $y_{\rm obs}$ and the results can be seen in Fig. \ref{mp:b}. 

Particularly, Fig. \ref{mp:b} shows that when there is an obstacle in the corridor ($box \ge 1$), the $y$ coordinate of MP will linearly increase with the rise of obstacle size. Meanwhile, through comparison, we found that the lateral distance from the MP to the edge of obstacle will always be about 0.4 meters despite the variation of obstacle size. Considering the average shoulder breadth of Japanese males aging from 20 to 24 is 0.45 m \cite{shoulder}, the empty space between pedestrians and the obstacle tend to be about 0.175 m. Therefore, we could presume that when evading an obstacle, a pedestrian would like to choose a MP that is near to the obstacle in order to get shorter walking distance but keep a short distance away from the obstacle in order not to collide with it. 

\subsection{Method to fit trajectory curves}\label{sec:gaussian:fitting}

Despite the MP could be easily calculated using the location information, it is hard to observe the SP and EP directly from the trajectories because real pedestrians cannot walk definitely straight and they tend to gradually change the walking direction other than abruptly. Therefore, we would first fit the trajectory curves as preparation for the estimation of SP and EP.   

\begin{figure}[ht]
\centering
\subfigure{
\includegraphics[width=0.45\textwidth]{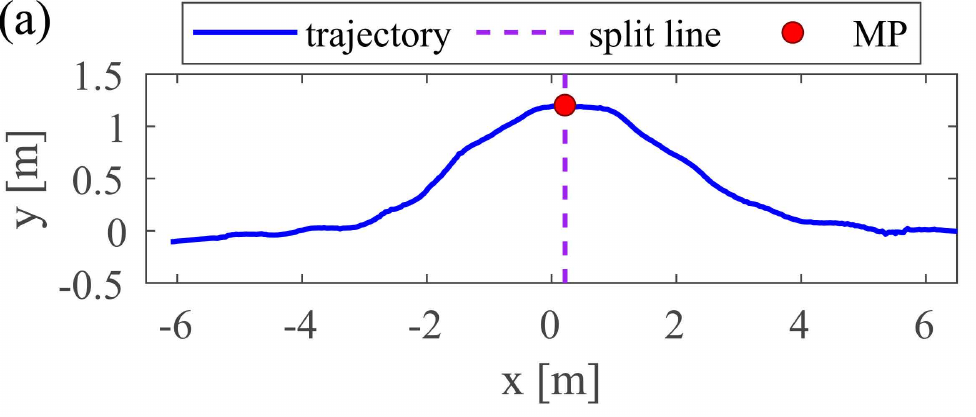}}
\hspace{1in} 
\subfigure{
\includegraphics[width=0.45\textwidth]{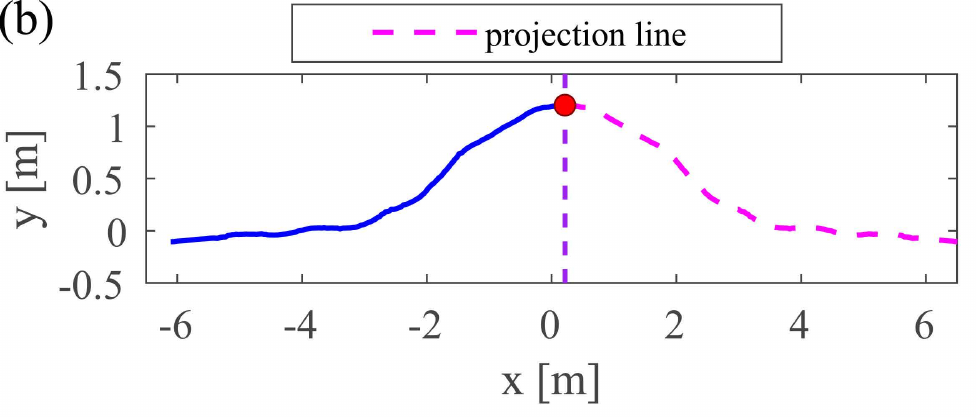}}
\subfigure{
\includegraphics[width=0.45\textwidth]{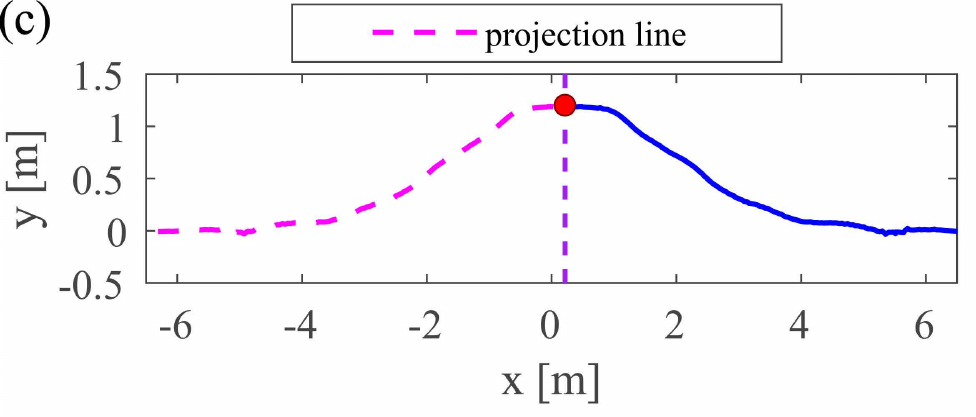}}
\hspace{1in} 
\subfigure{
\includegraphics[width=0.45\textwidth]{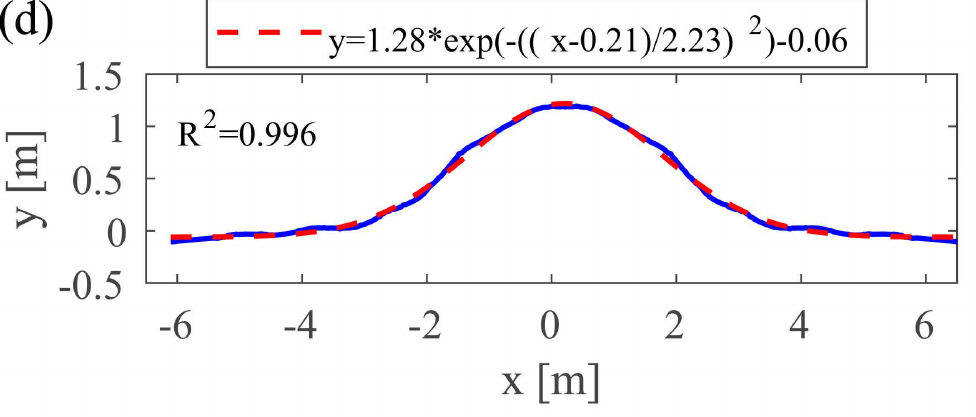}}
\subfigure{
\includegraphics[width=0.45\textwidth]{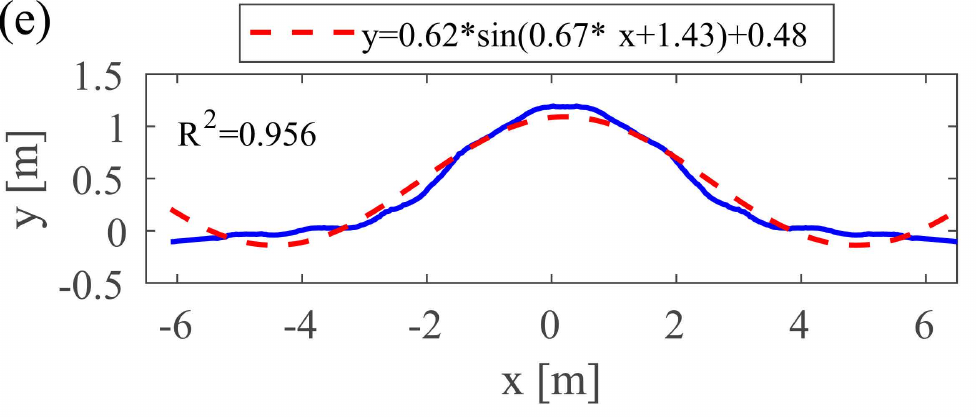}}
\hspace{1in} 
\caption{The fitting process of one trajectory curve. (a) take $x=x_{\rm m}$ as the split axis to separate one trajectory. (b) complement the left trajectory, i.e. the trajectory before the obstacle, through taking the projection line by the split axis. (c) complement the right trajectory, i.e. the trajectory after the obstacle, through taking the projection line by the split axis. (d) example of fitting the complemented trajectory in (b) with a Gaussian function. (e) example of fitting the complemented trajectory in (b) with a sine function.} 
\label{fitpro}
\end{figure}

In our experiments, a certain pedestrian will first set the MP as a temporal destination when evading the obstacle and set the exit as the next destination after passing by the obstacle. The two processes are separate and the trajectory before and after the MP are not symmetric. Therefore, we would separate the trajectory by the axis $x=x_{\rm m}$ and fit the two obtained trajectory curves respectively. For the convenience of fitting, each curve would be complemented through taking its projection line by the axis $x=x_{\rm m}$. This process can be seen in Fig. \ref{fitpro}(a)-\ref{fitpro}(c).  

To fit each separated trajectory curve, we would like to find a continuous derivatived function. According to the results of Fourier Transform in Sec. \ref{sec:common}, the main walking direction could be fitted by a trigonometric function. However, some pedestrians will first walk straight and then change his walking direction to evade the obstacle, which cannot be reproduced by a single trigonometric function. The only solution for this problem is building a piecewise function through combining the trigonometric function with another function. However, the continuity of derivative function at the piecewise points should be ensured and the parameters that need to be fitted would largely increase. Therefore, we would like to find another function that could reproduce both the evading process and the seemingly straight-walking process before the evading can be apparently observed. Based on the features of trajectory and slope curve, we would use a Gaussian function $f(x)$ as shown in Eq. \ref{eq:exp_gauss} to fit the walking trajectory. 
\begin{equation}
\label{eq:exp_gauss}
f(x)=A{\rm exp}\left( -\frac{(x-B)^2}{C^2}\right)+D
\end{equation}

In Eq. \ref{eq:exp_gauss}, $A$, $B$, $C$ and $D$ are constants that control the shape of the curves. $y=D$ represents the straight line a pedestrian tends to walk, $A$ represents the maximum deviation from the $y=D$, $B$ represents the $x$ coordinate of MP, i.e, $x_{\rm m}$, and $C$ represents the shape of the evading curve.

An comparison of fitting a certain trajectory curve using a Gaussian function and a trigonometric function can be seen in Fig. \ref{fitpro} (d)-\ref{fitpro}(e). The Gaussian function could fit both the seemingly straight-walking process and the evading process, while the sine function can only be used to fit the evading process.

Totally 128 pieces of trajectory data, i.e., the trajectories of all the 32 pedestrians when $box=1, 2, 3$ and $4$, with the body sway removed have been used for curve fitting. The coefficient of determination, i.e., $R^2$, is used to validate the accuracy of curve fitting. According to the fitting process shown in Fig. \ref{fitpro}, a trajectory curve will be cut into two parts by MP to fit the SP and EP respectively, which means each trajectory will be fitted for two times. As a result, totally 256 values of $R^2$ have been obtained and their values range from 0.907 to 0.999 with the average value being 0.989. By contrast, the corresponding values of $R^2$ when using a sine function range from 0.844 to 0.999 with the average value being 0.855, which is statistically lower than the results of Gaussian fitting (p-value $< 0.05$). Therefore, it could be proved that Gaussian function is better in fitting the evading trajectories of our experiments.

\subsection{Method to estimate SP and EP}\label{sec:gaussian:spepestimation}
The concept of confidence interval of Gaussian function has been widely used to show the concentration degree of all the samples and can be easily calculated. For instance, to a certain normal distribution function whose middle axis is $x=0$ and standard deviation is $\sigma$, respectively 68.27$\%$, 95.45$\%$ and 99.73$\%$ of the samples would be within the confidence region of $[-\sigma, \sigma]$, $[-2\sigma, 2\sigma]$ and $[-3\sigma, 3\sigma]$.

Analogically, as to a pedestrian who has to evade the obstacle, his trajectory would deviate from the desired horizontal axis. Therefore, we presume that major deviation of a pedestrian exists when his $x$ coordinate is among a certain confidence region, and assume the upper and lower boundary of the confidence region as SP and EP respectively. In the remaining part of this section, we would convert the trajectory curve into an equivalent normal distribution function and give a reasonable confidence region to estimate the locations of SP and EP.   

As has been mentioned that in Eq. \ref{eq:exp_gauss}, the function $f(x)$ could represent the deviation of the trajectory from the horizontal curve $y=D$. To obtain the confidence region, $f(x)$ is converted into a normal distribution function $\tilde{f}(x)$ whose integral value is one. 
\begin{equation}
\begin{split}
\label{eq:normdis}
\tilde{f}(x)=N(B,\sigma)=\frac{f(x)-D}{AC\sqrt{\pi}} \qquad \left(\sigma= \frac{C}{\sqrt{2}}\right)\\
\end{split}
\end{equation}

\begin{figure}[ht]
\centering
\includegraphics[width=0.5\textwidth]{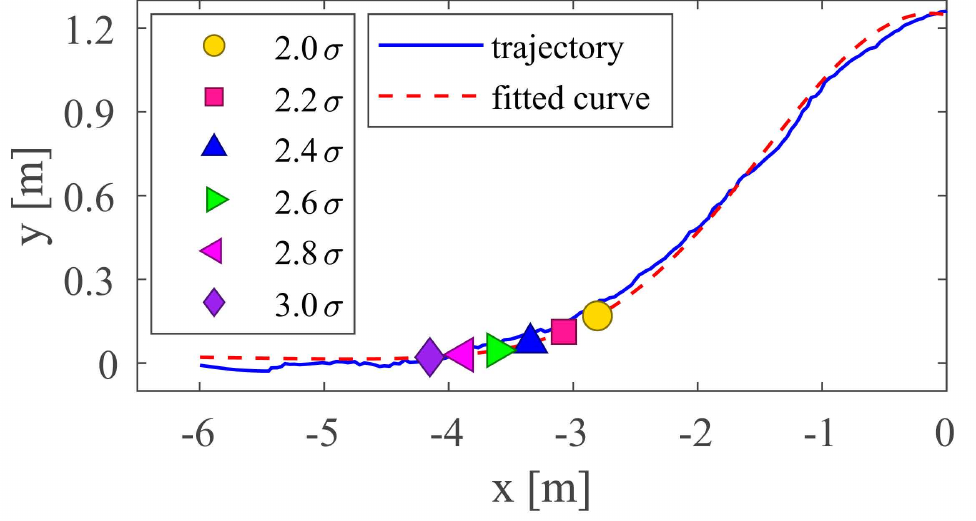}
\caption{Illustration of SP under different confidence intervals.} 
\label{interval}
\end{figure}

Through the deduction in Eq. \ref{eq:normdis}, it can be seen that the Gaussian function follows a normal distribution function $N(B,\sigma)$ whose average value equals to $B$ and standard deviation equals to $\sigma$. As mentioned before, the SP or EP could be considered as the boundary of a confidence interval. An Example of the SP under different confidence intervals can be seen in Fig. \ref{interval}. 

Theoretically, the curvature of the fitting trajectory depends on $\sigma$, and the confidence interval is linearly related to $\sigma$. As a result, any constants being the multiple of confidence interval can be used to represent the evading features of a certain pedestrian. For instance, when the confidence interval is $2\sigma$, it is reasonable to think that 95.45$\%$ deviation from the horizontal direction happens among the region of $x\in[-2\sigma+B, 2\sigma+B]$. Therefore, our next task is to find a reasonable confidence interval that is in accordance with our visual recognition. 

As mentioned in Sec. \ref{sec:gaussian:walking_trajectory}, pedestrians may behave `direct' and `indirect' when he has to evade. Therefore, we would qualitatively judge the number of `direct' and `indirect' pedestrians respectively before and after the obstacle, and the estimation results derived from a reasonable confidence interval should be in accordance with observation results. For better illustration, we define the number of `direct' pedestrians before and after the obstacle as $N\rm _{dir}^{sp}$ and $N{\rm _{dir}^{ep}}$, and the number of `indirect' pedestrians before and after the obstacle as $N\rm _{ind}^{sp}$ and $N\rm _{ind}^{ep}$. 

\begin{figure}[ht]
\centering
\subfigure[`direct' evading trajectory]{
\label{dir:a} 
\includegraphics[width=0.3\textwidth]{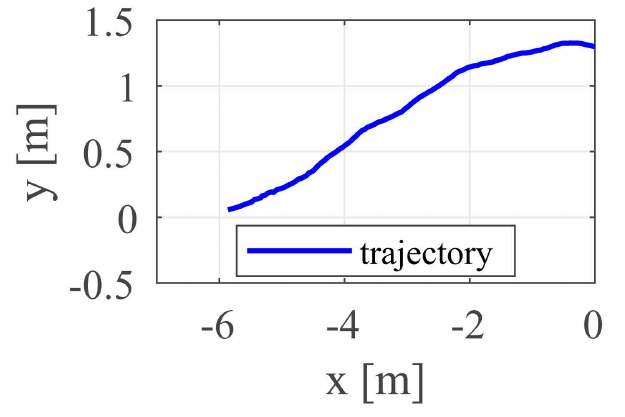}}
\subfigure[`indirect' evading trajectory]{
\label{dir:b} 
\includegraphics[width=0.3\textwidth]{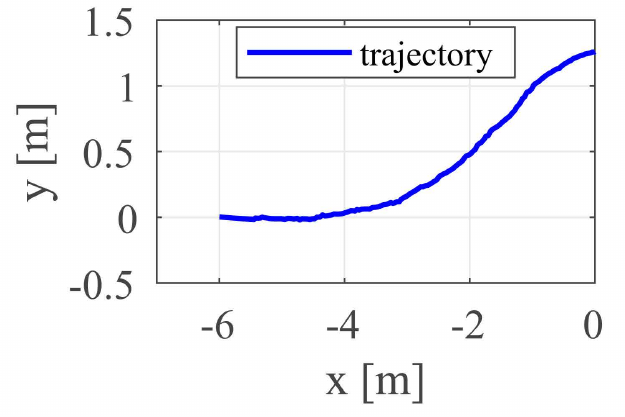}}
\subfigure[`unclear' evading trajectory]{
\label{dir:c} 
\includegraphics[width=0.3\textwidth]{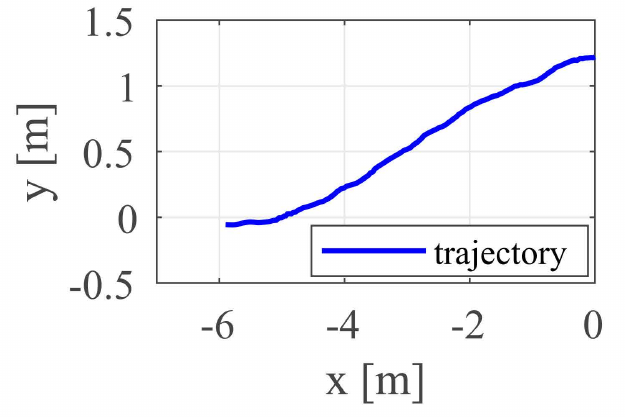}}
\caption{Examples of `direct', `indirect' and `unclear' pedestrian trajectories in qualitative observation.}
\label{dir} 
\end{figure}

Please note that the $x$ coordinate of the detection region is $[-6,6.5]$ m while the experimental region is  $[-7,7]$ m, which means we could not judge whether a pedestrian is `direct' or `indirect' if the SP or EP fall into the undetectable region. To distinguish observation results with theoretical definition, we define the number of `direct' evading, `indirect' evading, `direct' returning and `indirect' returning pedestrians in our qualitative observation as $N{\rm _{dir}^{sp}}'$, $N{\rm _{ind}^{sp}}'$, $N{\rm _{dir}^{ep}}'$ and $N{\rm _{ind}^{ep}}'$ respectively. To improve the reliability of manual observation, we only classify those with apparent behavior into the two mentioned types. If it is hard to confirm the classification, we would classify it into the type named `unclear', i.e. ${\it N}_{\rm unclear}^{\rm sp}$ and ${\it N}_{\rm unclear}^{\rm ep}$. Our classification is quite cautious, so the number of pedestrians belonging to `unclear' is not small. Examples of the three types can be seen in Fig. \ref{dir}. The results of qualitative observation can be seen in Tab. \ref{tab:observe}.

\begin{table}[ht] 
\centering
\caption{Results of qualitative observation classification of pedestrian trajectories.}
\label{tab:observe}
\begin{tabular}{|c|c c c|c c c|}
\hline
 & ${\it N}_{\rm dir}^{\rm sp'}$ & ${\it N}_{\rm ind}^{\rm sp'}$ & ${\it N}_{\rm unclear}^{\rm sp}$ & ${\it N}_{\rm dir}^{\rm ep'}$ & ${\it N}_{\rm ind}^{\rm ep'}$ &${\it N}_{\rm unclear}^{\rm ep}$\\
\hline
box=1&	9&	16&	7&	3&	18&	11 \\
box=2&	2&	20&	10&	4&	13&	14 \\
box=3&	9&	14&	9&	6&	11&	15 \\
box=4&	6&	12&	14&	7&	17&	8 \\
\hline
\end{tabular}
\end{table}

Please note that the `unclear' pedestrians in observation are possible to be either `direct' or `indirect', therefore the reasonable region of `direct' pedestrians can be obtained. On the other hand, the accurate estimated number of `direct' and `indirect' pedestrians in the detection region could be judged given a certain confidence interval. The validation results could be seen in Tab. \ref{tab:validate}, where cells with reasonable values are painted with yellow shadow. Comparison shows that the values corresponding to $2.5 \sigma$ are reasonable in all the cases, which means the estimation results under $2.5 \sigma$ is in accordance with the visual recognition. We hence emphasized the reasonable values with red and circled fonts and would use $2.5 \sigma$ as the confidence interval to estimate SP and EP hereinafter.


\begin{table}[ht]
\centering
\caption{Validation of fitting parameter through comparing estimation and manual observation results.}


    \begin{tabular}{|c|c|c|ccccc c ccccc|}
    \hline
   \multirow{2}[3]{*}{\textit{N}} & \multirow{2}[3]{*}{\textit{box}} & \multicolumn{1}{c|}{reasonable}& \multicolumn{11}{c|}{confidence interval} \\
\cline{4-14}        &       &   range    & 2.0$\sigma$ & 2.1$\sigma$ & 2.2$\sigma$ & 2.3$\sigma$ & \multicolumn{1}{c}{2.4$\sigma$} & \multicolumn{1}{c}{2.5$\sigma$} & 2.6$\sigma$ & 2.7$\sigma$ & 2.8$\sigma$ & 2.9$\sigma$ & 3.0$\sigma$\\
\hline
\cline{9-9}   
    \multirow{4}[1]{*}{$N_{\rm dir}^{\rm sp'}$} & 1     & [9, 16]  & 5     & 5     & 6     & 6     & 7     & \cellcolor[rgb]{1,1,0}\textcolor[rgb]{ 1,  0,  0}{\textbf{\Large{\textcircled{\small{12}}}}} & \cellcolor[rgb]{ 1,  1,  0}13 & \cellcolor[rgb]{ 1,  1,  0}13 & \cellcolor[rgb]{ 1,  1,  0}15 & \cellcolor[rgb]{ 1,  1,  0}15 & \cellcolor[rgb]{ 1,  1,  0}15 \\
          & 2     & [2, 12]  & \cellcolor[rgb]{ 1,  1,  0}2 & \cellcolor[rgb]{ 1,  1,  0}3 & \cellcolor[rgb]{ 1,  1,  0}7 & \cellcolor[rgb]{ 1,  1,  0}9 & \cellcolor[rgb]{ 1,  1,  0}11 & \cellcolor[rgb]{ 1,  1,  0}\textcolor[rgb]{ 1,  0,  0}{\textbf{\Large{\textcircled{\small{11}}}}} & 13    & 15    & 16    & 16    & 17 \\
          & 3     & [9, 18]  & \cellcolor[rgb]{ 1,  1,  0}9 & \cellcolor[rgb]{ 1,  1,  0}11 & \cellcolor[rgb]{ 1,  1,  0}12 & \cellcolor[rgb]{ 1,  1,  0}13 & \cellcolor[rgb]{ 1,  1,  0}14 & \cellcolor[rgb]{ 1,  1,  0}\textcolor[rgb]{ 1,  0,  0}{\textbf{\Large{\textcircled{\small{16}}}}} & 19    & 19    & 20    & 23    & 24 \\
          & 4     & [6, 20]  & \cellcolor[rgb]{ 1,  1,  0}8 & \cellcolor[rgb]{ 1,  1,  0}9 & \cellcolor[rgb]{ 1,  1,  0}11 & \cellcolor[rgb]{ 1,  1,  0}14 & \cellcolor[rgb]{ 1,  1,  0}15 & \cellcolor[rgb]{ 1,  1,  0}\textcolor[rgb]{ 1,  0,  0}{\textbf{\Large{\textcircled{\small{17}}}}} & \cellcolor[rgb]{ 1,  1,  0}17 & \cellcolor[rgb]{ 1,  1,  0}19 & 22    & 23    & 25 \\
    \hline
    \multirow{4}[1]{*}{$N_{\rm dir}^{\rm ep'}$} & 1     & [3, 14]  & \cellcolor[rgb]{ 1,  1,  0}3 & \cellcolor[rgb]{ 1,  1,  0}5 & \cellcolor[rgb]{ 1,  1,  0}5 & \cellcolor[rgb]{ 1,  1,  0}8 & \cellcolor[rgb]{ 1,  1,  0}10 & \cellcolor[rgb]{ 1,  1,  0}\textcolor[rgb]{ 1,  0,  0}{\textbf{\Large{\textcircled{\small{11}}}}} & \cellcolor[rgb]{ 1,  1,  0}12 & \cellcolor[rgb]{ 1,  1,  0}12 & \cellcolor[rgb]{ 1,  1,  0}12 & \cellcolor[rgb]{ 1,  1,  0}12 & \cellcolor[rgb]{ 1,  1,  0}12 \\
          & 2     & [4, 18]  & 3     & \cellcolor[rgb]{ 1,  1,  0}5 & \cellcolor[rgb]{ 1,  1,  0}6 & \cellcolor[rgb]{ 1,  1,  0}7 & \cellcolor[rgb]{ 1,  1,  0}10 & \cellcolor[rgb]{ 1,  1,  0}\textcolor[rgb]{ 1,  0,  0}{\textbf{\Large{\textcircled{\small{10}}}}} & \cellcolor[rgb]{ 1,  1,  0}13 & \cellcolor[rgb]{ 1,  1,  0}15 & \cellcolor[rgb]{ 1,  1,  0}18 & 19    & 20 \\
          & 3     & [6, 21] & 2     & 2     & 3     & 4     & \cellcolor[rgb]{ 1,  1,  0}6 & \cellcolor[rgb]{ 1,  1,  0}\textcolor[rgb]{ 1,  0,  0}{\textbf{\Large{\textcircled{\small{9}}}}} & \cellcolor[rgb]{ 1,  1,  0}10 & \cellcolor[rgb]{ 1,  1,  0}12 & \cellcolor[rgb]{ 1,  1,  0}13 & \cellcolor[rgb]{ 1,  1,  0}15 & \cellcolor[rgb]{ 1,  1,  0}18 \\
          & 4     & [7, 15]  & 3     & 5     & \cellcolor[rgb]{ 1,  1,  0}7 & \cellcolor[rgb]{ 1,  1,  0}9 & \cellcolor[rgb]{ 1,  1,  0}9 & \cellcolor[rgb]{ 1,  1,  0}\textcolor[rgb]{ 1,  0,  0}{\textbf{\Large{\textcircled{\small{10}}}}} & \cellcolor[rgb]{ 1,  1,  0}11 & \cellcolor[rgb]{ 1,  1,  0}13 & 16    & 18    & 19 \\
    \hline
    \end{tabular}%

\label{tab:validate}
\end{table}

\begin{figure}[ht]
\centering
\subfigure[`indirect' evading and returning]{
\label{sme_exp:a} 
\includegraphics[width=0.45\textwidth]{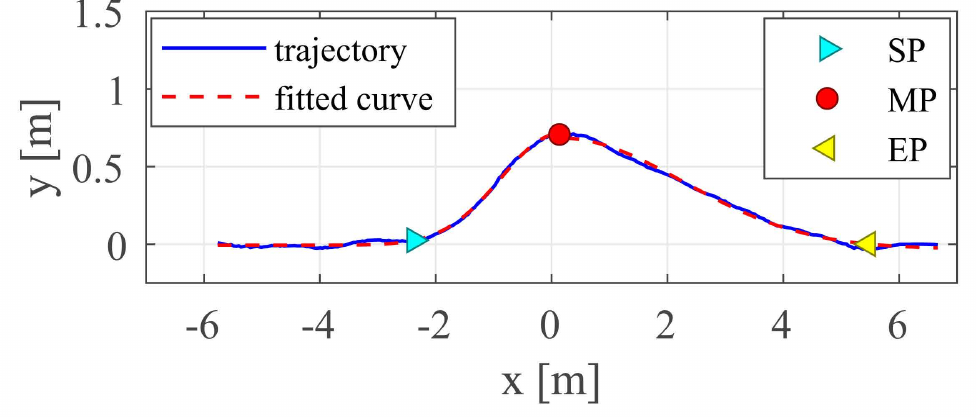}}
\subfigure[`indirect' evading but `direct' returning]{
\label{sme_exp:b} 
\includegraphics[width=0.45\textwidth]{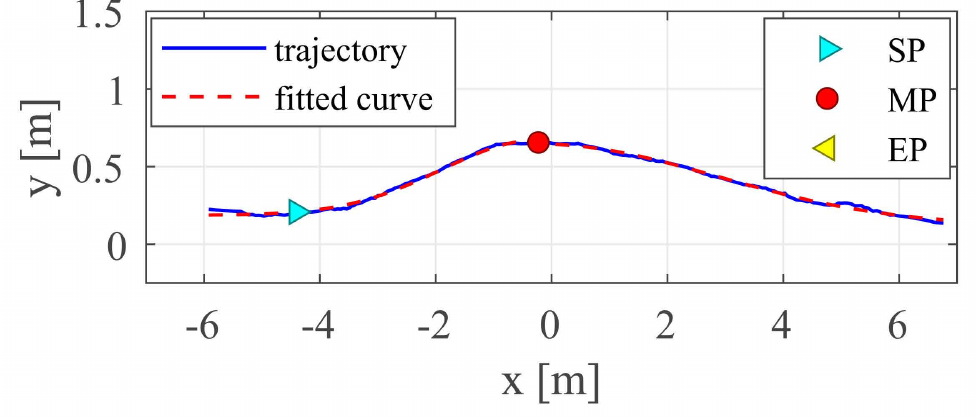}}
\hspace{1in} 
\subfigure[`direct' evading but `indirect' returning]{
\label{sme_exp:c} 
\includegraphics[width=0.45\textwidth]{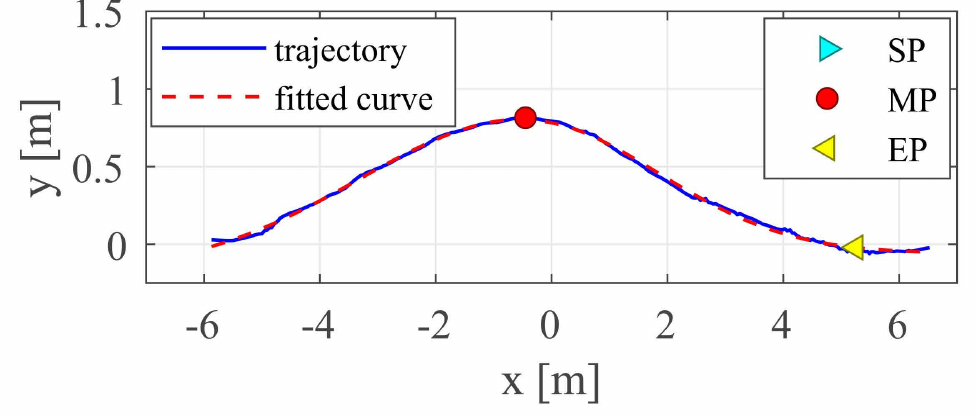}}
\subfigure[`direct' evading and returning]{
\label{sme_exp:d} 
\includegraphics[width=0.45\textwidth]{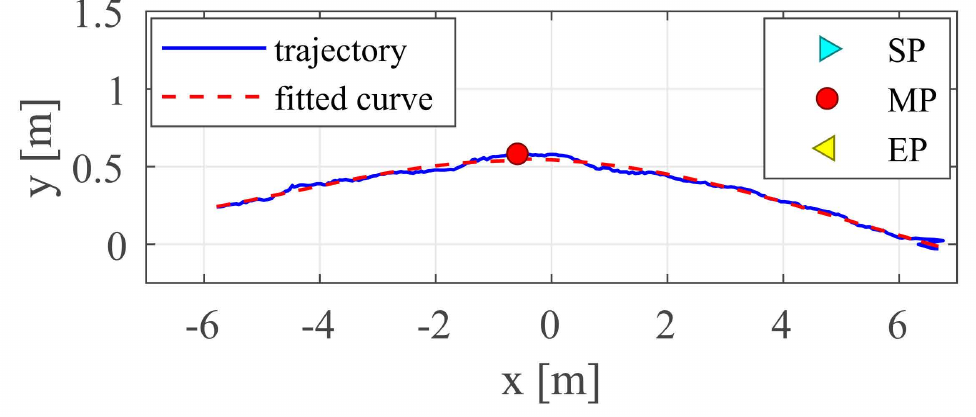}}
\caption{Examples of curve fitting and three critical points.}
\label{sme_exp} 
\end{figure}

Based on the curve fitting method in Sec. \ref{sec:gaussian:fitting} and the reasonable confidence region in this subsection, all the 128 trajectories could be fitted and the three critical points, i.e. SP, MP and EP, could be obtained. Four examples have been chosen to illustrate the outcome of the estimation, which can be seen in Fig. \ref{sme_exp}. 

\subsection{Variation of SP and EP}\label{sec:gaussian:distribution}

After obtaining the coordinates of SP and EP, the influence of obstacle width on the SP and EP could be analyzed. 
To better illustrate, we assume the coordinates of SP as $[x_{\rm s}, y_{\rm s}]$ and EP as $[x_{\rm e},y_{\rm e}]$. For those `direct' pedestrians before or after the obstacle, we presume $x_{\rm s}=-7$ m or $x_{\rm e}=7$ m because $7$ m is the distance from the obstacle to the entrance/exit. Therefore, the variation of $x_{\rm s}$ and $x_{\rm e}$ as well as their corresponding one standard deviation $\sigma_s$ and $\sigma_e$ with the increase of obstacle size could be obtained and shown in Tab. \ref{tab:avgxse}. 
\begin{table}[ht]
\centering
\caption{Variation of average $x$ coordinate of SP and EP with obstacle size.}
\label{tab:avgxse}
\begin{tabular}{|c|c|c|c|c|}
\hline
$box$  & $|x_{\rm s}|$ & $x_{\rm e}$& $\sigma_s$ & $\sigma_e$ \\
 \hline
1 &	5.40 &	5.47  &	1.33 & 1.24\\
\hline
2 &	5.34 & 5.67 &	1.41 & 1.15\\
\hline
3 &	5.73 &	5.84 &	1.28 & 0.83\\
\hline
4 & 5.93 & 5.46 &	1.13 & 1.10\\
\hline
\end{tabular}
\end{table}


Tab. \ref{tab:avgxse} shows that $|x_{\rm s}|$ will basically increase with the rise of obstacle size, which means pedestrians tend to evade earlier if the obstacle becomes larger. In our experiments, the trajectories of the same pedestrian under different obstacle sizes can be obtained because the pedestrians have all been numbered. Therefore, we could use the paired-samples t-test method to check whether the influence of obstacle size is statistically apparent. 

Results show that when $box\leq 3$, the influence of obstacle size on SP is not apparent (the p-value of any two conditions larger than 0.05). When $box= 4$, the $|x_{\rm s}|$ apparently increases compared with the conditions when $box= 1$ and $box= 2$. Therefore, we presume that when the obstacle size is small ($box\leq 2$), pedestrians are mainly affected by the walking habits or inertia other than the obstacle size. Nevertheless, when the obstacle size becomes larger ($box\geq3$),  pedestrians become more urgent to evade the obstacle because they have realized that the increase of obstacle size would affect their walkable space. As a result, when $box= 3$, $|x_{\rm s}|$ becomes larger but the variation is not apparent, which means pedestrians are  slightly affected by the obstacle size but the influence is not dominating. The urgency of pedestrians becomes the maximum when $box= 4$, under which condition the obstacle size is the main contributing factor that affects the evading behavior. Therefore, we presume that the obstacle will have apparent influence on the location of SP only when the obstacle size is large enough to become the main contributing factor. 

By contrast, the influence of obstacle size on the location of EP is not apparent (the p-value of any two conditions larger than 0.05). We presume that the reason is pedestrians will not feel urgent after passing by the obstacle, making their movement mainly decided by their walking tendency to whether return to the middle horizontal axis or not.

\section{Analysis on the evading preference}
Participants have certain preference to behave `direct' or `indirect' during the evading process. Meanwhile, they have preference in choosing whether to turn left or right. In this section, we would like to discuss the two kinds of preferences and their mutual relation. The psychological motivations behind would also be analyzed.

\subsection{Preference of `direct' and `indirect' behavior}

The number of `direct' and `indirect' pedestrians in the experiment region, i.e. $x\in [-7,\ 7]$ m, have been estimated as shown in Tab. \ref{tab:finalresult} according to their definitions in Sec. \ref{sec:gaussian:spepestimation}. Please note that the numbers of `direct' pedestrians in Tab. \ref{tab:finalresult} are generally smaller than the corresponding values in Tab. \ref{tab:validate}. The reason is that Tab. \ref{tab:validate} is estimated based on the detection region, i.e. $[-6,\ 6.5]$ m. Considering the detection region is smaller than the experiment region, some pedestrians that are `direct' in Tab. \ref{tab:validate} may become `indirect' in Tab. \ref{tab:finalresult}.  

\begin{table}[ht]
\centering
\caption{Final results of SP and EP estimation.}
\label{tab:finalresult}
\begin{tabular}{|c|c c|c c|}
\hline
 box & $N\rm _{dir}^{sp}$ & $N\rm _{ind}^{sp}$ & $N{\rm _{dir}^{ep}}$&$N\rm _{ind}^{ep}$\\
\hline
1& 9 & 23 & 9 & 23 \\
2& 9& 23 & 7 & 25\\
3& 10 & 22 & 5 & 27 \\
4& 12 & 20 & 7 & 25\\
\hline
\end{tabular}
\end{table}

We presume that the `direct' and `indirect' behavior both have their reasonable psychological motivation. Tab. \ref{tab:finalresult} shows that the number of `indirect' pedestrians is larger than that of `direct' pedestrians both before and after passing by the obstacle, which indicates that the majority of pedestrians tend to first walk straight for some distance and then begin to change the walking direction. Nevertheless, the behavior of `direct' pedestrians should be more effective because their walking distance is the shortest. 

\begin{figure}[ht]
\centering
\subfigure{
\label{lefvelo:a} 
\includegraphics[width=0.4\textwidth]{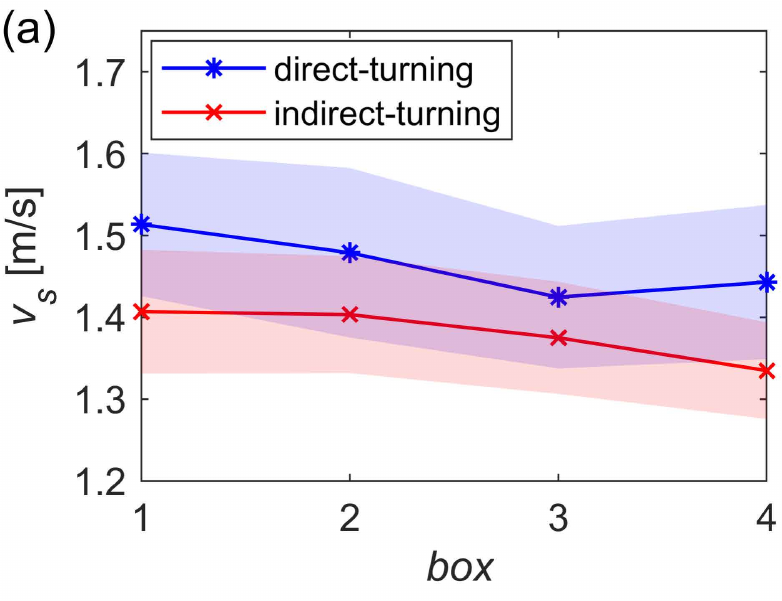}  }
\qquad
\subfigure{
\label{lefvelo:b} 
\includegraphics[width=0.4\textwidth]{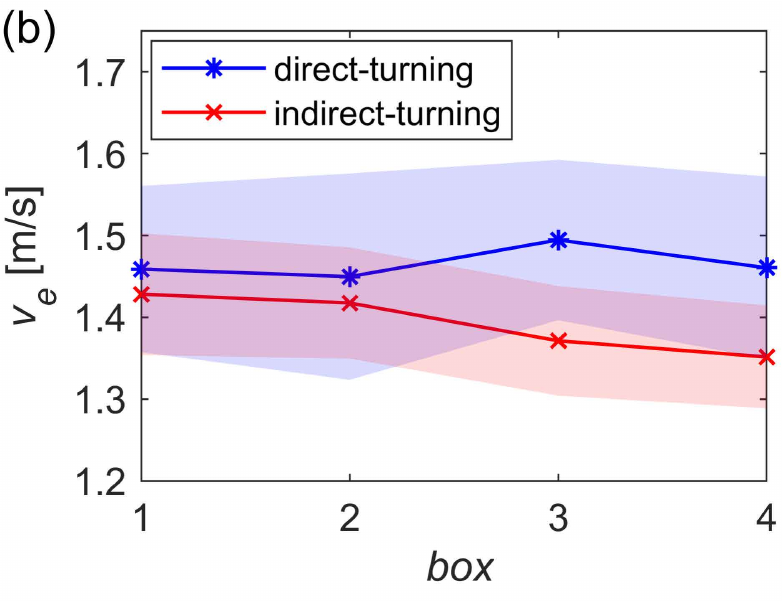}}
\caption{Comparison of average velocity between `direct' and `indirect' pedestrians.}
\label{lefvelo} 
\end{figure}

Therefore, we presume that psychological motivation of `direct' pedestrians and `indirect' are different. `Direct' pedestrians are urgent to leave the corridor and hence choose the shortest route. `Indirect' pedestrians tend to first walk in a straight line for some time because of the inertia or other reasons, and then take actions to evade the obstacle because they realize some sense of urgency to avoid collision when coming close enough to the obstacle. In other words, `direct' would like to traverse the corridor in the most efficient way, while `indirect' pedestrians may prefer to walk at their desired or the most comfortable way. This presumption can be strongly supported by the fact that `direct' pedestrians tend to walk faster than `indirect' pedestrians as shown in Fig. \ref{lefvelo}, where $v_s$ and $v_e$ respectively represents the walking velocity before and after the obstacle.




 Fig. \ref{lefvelo} shows that the velocity of `direct' pedestrians tends to be higher than that of `indirect' pedestrians despite the increase of obstacle width. To statistically validate the velocity difference in Fig. \ref{lefvelo}, we first use t-test to check whether the obstacle size has influence on walking velocity. Through calculating the p-value of any two conditions with different obstacle sizes, no clear relation has been found between the obstacle size and the walking velocity (all the p-value larger than 0.05). Therefore, we divide the trajectories both before and after the obstacle into direct-turning and indirect-turning groups regardless of obstacle width. Using t-test to check whether the velocity corresponding to the two groups are apparently different, we have found that the velocity of `direct' pedestrians tend to be higher than `indirect' pedestrians both before and after the obstacle (both p-value smaller than 0.05). 


\subsection{Preference of left and right turning behavior}
Despite our experimental scenario is symmetric, pedestrians may have their preference of whether to turn to the left or right before passing by the obstacle. For better illustration, we define the total pedestrian number as $N$, the number of left-turning pedestrians as $L$, and the number of right-turning pedestrians as $R$. Their values can be seen in Tab. \ref{tab:turning}.

\begin{table}[ht]
\centering
\caption{Number of left and right turning pedestrians and the corresponding ratio.}
\label{tab:turning}
\begin{tabular}{|c|c|c|c|}
\hline
 $box$&$L$&$R$&$L/N$\\
 \hline
1&20&12&0.63\\
\hline
2&	20&	12&	0.63\\
\hline
3&	19&	13&	0.59\\
\hline
4&	17&	15&	0.53\\
\hline
\end{tabular}
\end{table}

In our experiments, the number of pedestrians getting into the corridor from the back of the upper wall and the bottom wall are the same. Nevertheless, it can be seen in Tab. \ref{tab:turning} that the number of left-turning pedestrians is always larger than that of right-turning ones. As a result, we could presume that our participants have the tendency to turn left before the obstacle, which may have some relation with the left-side driving rules and the left-side passing in subway stations in Japan.

\subsection{Relation between direct-indirect and left-right preferences}



Among the left and right turning pedestrians, the number of those who are `direct' and `indirect' before the obstacle can be seen in Tab. \ref{tab:dirturn}. From the perspective of left-right turning, more pedestrians tend to turn left. From the perspective of direct-indirect turning, the number of `indirect' evading pedestrians is over two times larger than `direct' pedestrians. On the other hand, for left-turning pedestrians, most of them tend to behave `indirect'. However, for right-turning pedestrians, the number of `direct' and `indirect' pedestrians are similar. Therefore, we presume that right turning pedestrians have a stronger tendency to behave `direct' compared with left-turning pedestrians. 

\begin{table}[ht]
\centering
\caption{Number of left-right and direct-indirect turning pedestrians.}
\label{tab:dirturn}
\begin{tabular}{|c|c|c|c|c|}
\hline
 & $L$ & $R$ & total\\
    \hline
 $N{\rm _{dir}^{sp}}$ &  16 & 24& 40\\
    \hline
 $N{\rm _{ind}^{sp}}$ &  60 & 28 & 88\\ 
    \hline
 total & 76 & 52 & 128 \\
    \hline
\end{tabular}
\end{table}

\begin{figure}[ht]
\centering
\includegraphics[width=0.5\textwidth]{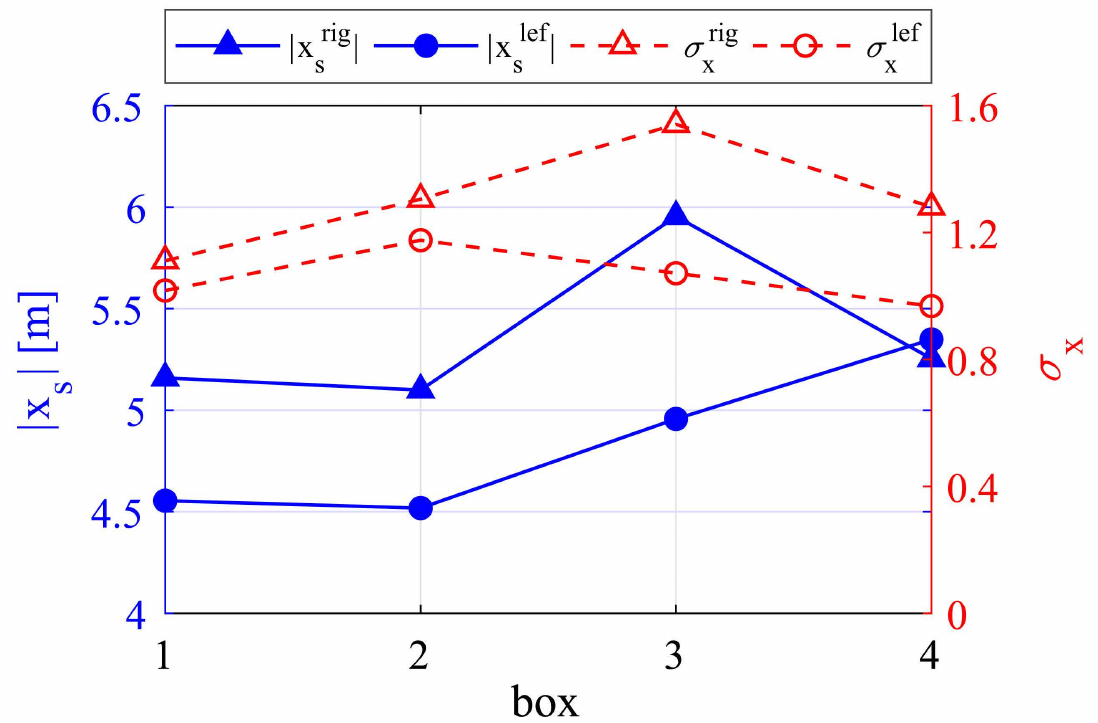}
\caption{Variation of average $|x_{\rm s}|$ for left and right turning `indirect' pedestrians.}
\label{spxleri}
\end{figure}

The results of $\chi^2$ test could support this presumption with the corresponding p-value being 0.0026 (smaller than 0.05). As a result, we could prove that pedestrians who turn right statistically have a stronger tendency to directly change the walking direction when getting into the entrance. We presume one reason is that pedestrians are more used to walk by the left side, making them tend to evade earlier when choosing to walk by the right side which they are relatively not used to. This presumption can be supported by the fact that for `indirect' pedestrians, right-turning pedestrians will begin to evade the obstacle earlier than left-turning ones as shown in Fig. \ref{spxleri}.


Fig. \ref{spxleri} shows the average value of $|x_{\rm s}|$ and standard deviation for those left-turning and right-turning pedestrians among the `indirect' pedestrians. $|x_{\rm s}^{\rm rig} |$ represents the average $|x_{\rm s}|$ of right-turning pedestrians,  $|x_{\rm s}^{\rm lef} |$ represents the average $|x_{\rm s}|$ of left-turning pedestrians, $ \sigma_x^{\rm rig}$ represents the one standard deviation of all the $|x_{\rm s}|$ of right-turning pedestrians and $ \sigma_x^{\rm lef}$ represents the one standard deviation of all the $|x_{\rm s}|$ of left-turning pedestrians.

Fig. \ref{spxleri} shows that when $box\leq 3$, those right-turning pedestrians tend to begin to evade the obstacle at a location more than 0.5 meters farther from the obstacle than left-turning pedestrians. When $box=4$, the $|x_{\rm s}|$ of left-turning and right-turning pedestrians are nearly the same. Therefore, we presume that the right-turning pedestrians tend to begin to evade the obstacle earlier than the left-turning pedestrians when the obstacle size is not the main contributing factor during the evading process, i.e., $box\leq 3$ as shown in Sec. \ref{sec:gaussian:distribution}. For statistical illustration, we first use t-test to verify that the obstacle size would not apparently affect the $|x_{\rm s}|$ because the p-value of any two conditions that meets $box\leq 3$ are smaller than 0.05. Therefore, we consider the data of the three obstacle sizes as the same data set and compare the $|x_{\rm s}|$ of left-turning and right-turning pedestrians. The  p-value obtained from t-test is $0.027(<0.05$), which proves that the right-turning `indirect' pedestrians will evade earlier than the left-turning `indirect' pedestrians.

\section{Conclusion}
A thorough analysis of the evading behavior of individual pedestrian when confronting with an obstacle has been conducted. The obstacle width would be adjusted to explore the influence in the experiments, and pedestrians have to change their walking directions in order to traverse the corridor. Fourier Transform has been used to analyze the slope curves derived from trajectories. Results show that pedestrian movement is mainly composed of two parts: body sway and variation of walking direction. 

The body sway has relatively permanent frequency and amplitude that would not be affected by the obstacle. Accordingly, the gait features of pedestrians including sway amplitude, frequency, stride length and speed could all be calculated and their mutual relation could be obtained. Results show that higher walking speed usually corresponds to longer stride length and higher frequency. Nevertheless, the speed mainly depends on the stride length rather than the frequency. Meanwhile, higher stride frequency usually corresponds to shorter stride length and smaller body sway amplitude.


The walking direction is mainly featured by the three critical points, i.e. SP, MP and EP. The Gaussian function has been used to fit the walking trajectories. Accordingly, the critical points has been estimated and validated with the help of observation results. The features of MP show that pedestrians tend to laterally pass a location about 0.4 m away from the obstacle despite the obstacle size. 
Besides, the location of SP is mainly affected by the obstacle when the obstacle size is large enough to make pedestrians feel urgent to evade. Nevertheless, the location of EP is not affected by the obstacle size, which we presume is because pedestrians do not feel urgent after passing by the obstacle. 

In addition, the direct-indirect evading and left-right turning preference of participants have been illustrated. `Direct' pedestrians tend to pass by the corridor more efficiently with shorter route distance and higher walking velocity compared with `indirect' pedestrians. Besides, pedestrians have a left-turning preference and tend to be more prepared when turning right. As a result, right-turning pedestrians have a stronger tendency to behave `direct' or evade earlier if they behave 'indirect' before evading the obstacle.

We presume the results of this paper could help exploring the essence of evading behavior and contribute to the relative modeling of pedestrian dynamics under the influence of obstacles. 

\section{Acknowledgements}
This work was supported by JST-Mirai Program Grant Number JPMJMI17D4, Japan and JSPS KAKENHI Grant Number 15K17583. The first author is financially supported by Chinese Scholarship Council. In addition, the authors would like to thank other team members of Nishinari group for the assistance during the planning and conduction of the experiments in this paper.



\end{document}